\documentclass[acmsmall,screen,nonacm]{acmart}

\AtBeginDocument{%
  \providecommand\BibTeX{{%
    \normalfont B\kern-0.5em{\scshape i\kern-0.25em b}\kern-0.8em\TeX}}}

\usepackage[linesnumbered,ruled,vlined]{algorithm2e}
\usepackage{amsmath,amsfonts}
\usepackage{textcomp}
\usepackage{xcolor}
\usepackage{caption}
\usepackage{subcaption}
\usepackage{multirow}
\usepackage{graphicx}
\usepackage{booktabs}
\usepackage{makecell}
\usepackage{minted}
\usepackage{enumitem}
\usepackage{listings}
\usepackage{tcolorbox}
\usepackage{multicol} 
\usepackage{subcaption}
\usepackage{ulem}
\usepackage{wrapfig}
\usepackage{rotating} 
\usepackage[noend]{algpseudocode}
\usepackage{amsmath}
\usepackage{hyperref}

\usepackage{wrapfig}
\usepackage{subcaption}
\usepackage{graphicx}

\setminted{fontsize=\small,
    xleftmargin=6mm,
   linenos,
   frame=lines,
   breaklines}
\begin{document}

\newcommand{\todo}[1]{\textbf{\color{red}{Ravishka: #1} }}

\newcommand{\wei}[1]{\textcolor{blue}{Wei: [#1]}}

\newcommand{\ww}[1]{\textcolor{blue}{Weihang: [#1]}}

\newcommand{\td}[1]{\textcolor{blue}{ToDo: [#1]}}

\newcommand{\zijie}[1]{{\color{purple}[Zijie: #1]}}

\newcommand{\zeqing}[1]{{\textcolor{blue}[Zeqing: #1]}}

\newcommand{\jiajun}[1]{{\color{cyan}[Jiang: #1]}}

\title{Characterizing Overconfident Failure in LLM-Based Code Generation}

\author{Ravishka Rathnasuriya}
\orcid{0009-0005-6129-2865}
\email{ravishka.rathnasuriya@utdallas.edu}
\affiliation{%
  \institution{University of Texas at Dallas}
  \city{Richardson}
  \country{USA}
}

\author{Wei Yang}
\orcid{0000-0002-5338-7347}
\email{yang\_wei@fudan.edu.cn}
\affiliation{%
  \department{Institute of Systems for Advanced Computing}
  \institution{Fudan University}
  \city{Shanghai}
  \country{China}
}


\begin{abstract}
Large language models (LLMs) are increasingly used for automated code generation, but generated programs can appear syntactically plausible while still failing execution-based correctness checks. Existing validation methods, such as testing and program analysis, remain essential but are often incomplete, costly, or applied only after generation. Model-derived uncertainty is therefore a natural early reliability signal.

This paper studies the \emph{dilemma of overconfidence} in code LLMs where incorrect programs are often generated with token-level confidence comparable to correct programs. We study this dilemma across four open-source code models and three execution-based benchmarks. Our analysis begins by investigating whether existing uncertainty metrics provide reliable proxies for execution correctness in code generation. We then characterize overconfidence at both global and local token levels, asking whether incorrect programs remain indistinguishable from correct ones under confidence and entropy summaries, including selective generation and the limits of instruction tuning. Finally, we evaluate whether common mitigation strategies reduce this failure mode, and we examine latent representations as exploratory evidence for future reliability mechanisms.

Our study yields four findings. First, existing uncertainty signals provide only partial and model-dependent evidence of execution failure. Second, overconfidence persists at both program and token levels, and uncertainty-based selection does not consistently improve accepted-set accuracy. Third, instruction tuning can increase certainty on failing generations without consistently improving correctness discrimination. Fourth, common mitigation techniques improve specific aspects of reliability but do not reliably resolve overconfident failure. Our exploratory latent analysis suggests that hidden representations may encode correctness-related signals that output confidence does not expose. Together, these findings identify overconfident failure as a software-engineering reliability problem and offer a practical path toward building more reliable LLM-driven code-generation systems.

\end{abstract}
\maketitle

\section{Introduction}

Large language models (LLMs) have become increasingly effective at code generation, synthesizing implementations from natural-language specifications, programming tasks, and partial code contexts~\cite{roziere2023code,guo2024deepseek,jiang2024self,evalplus,evalperf}. They are now integrated into software development workflows to assist with implementation, debugging, testing, and repair~\cite{github_copilot,openai_codex,deepmind_alphaevolve,deepmind_alphacode}. However, generated code is not self-validating. A program may be syntactically valid, idiomatic, and plausible while still violating the intended specification, especially on edge cases revealed only through execution. Thus, a central reliability challenge for LLM-based software engineering is not only whether a model can generate code, but whether a system can determine when generated code should be trusted.

Existing validation approaches primarily rely on execution-based evaluation~\cite{li2024doce,peng2025perfcodegen,ni2023lever,tian2025fixing} or program analysis~\cite{yao2025training,shaikhelislamov2024codepatchllm,chapman2024interleaving}. These techniques are essential, but they do not fully solve the deployment-time reliability problem. Execution requires tests, input generators, or behavioral oracles, which may be incomplete, expensive, or unavailable for each generated candidate. Program analysis can identify specific defect classes, but it is often conservative, property-specific, and does not generally estimate whether an arbitrary generated program satisfies its specification. Moreover, both execution and analysis are typically applied after generation. This limits their ability to provide lightweight, model-intrinsic reliability signals during interactive code generation.

A natural alternative is an \textit{\textbf{uncertainty-aware}} code-generation system. In an ideal setting, the system would know whether a generated program generalizes to unseen tests or real user inputs. Such a generalizability oracle is unavailable in practice, because functional correctness can only be established through validation against a specification, tests, or downstream behavior. Therefore practical systems must rely on \textit{proxies}. The broader machine-learning literature operationalizes this proxy through uncertainty metrics derived from predictive distributions, confidence scores, entropy, margins, calibration methods, and related quantities. Such metrics have been extensively studied in classification tasks~\cite{guo2017calibration,wang2020dissector,hendrycks2018baseline,gal2016dropout,alon2019code2vec,xiao2019quantifying,vasudevan2019towards,corbiere2019addressing,monarch2021human,steinhardt2016unsupervised,shannon1948mathematical} and more recently in natural language generation (NLG)~\cite{vashurin2024benchmarking,xie2025empirical,xia2025survey,bakman2025reconsidering,ielanskyi2025addressing,huang2025look}, where they are used as proxies for output correctness, factuality, or risk. These estimates also support \textit{\textbf{selective abstention}}~\cite{xin2021art,calder1999selective,gu2023evaluation}, where high-risk outputs are withheld or routed to additional validation.

However, uncertainty is less directly related to correctness in code generation than in classification or open-ended NLG. In classification, uncertainty is defined over the label distribution, and evaluation is performed on the predicted label. In open-ended NLG, uncertainty is commonly estimated over the generated response and evaluated against response-level properties such as correctness or factuality. In code generation, model probabilities are assigned during token generation, whereas correctness is determined by the execution behavior of the completed program. Therefore, a generated program can have high token likelihood and still fail to satisfy its specification. Thus, token- and sequence-level uncertainty are only \textit{proxies} for execution correctness.

Using token- and sequence-level uncertainty as proxies for execution correctness creates a \emph{dilemma of overconfidence} in code generation. A generated program may receive high model confidence and still fail execution-based evaluation. This mismatch is consistent with how code LLMs are trained and evaluated. First, autoregressive training optimizes next-token prediction rather than whether model confidence reflects execution correctness. Second, standard code-generation evaluation measures generation success rather than the reliability of model confidence. Pass@1 and pass@k measure~\cite{li2024doce,peng2025perfcodegen,ni2023lever,tian2025fixing} whether generated programs pass the evaluation tests, whereas binary grading gives no credit to abstention~\cite{kalai2025language}. Therefore, such evaluation  does not assess whether failing generations receive lower confidence than passing ones. Testing and program analysis provide evidence more directly tied to execution correctness, but they require external validation and may be costly or incomplete. Therefore, model-derived uncertainty  remains attractive as an early reliability signal, despite the possibility of overconfident failures.


There is growing research interest in using uncertainty to support reliable code generation. Existing studies pursue three main directions. First, they evaluate uncertainty estimators as predictors of code correctness~\cite{huang2025look,vasconcelos2022generation}. Second, they calibrate confidence against observed correctness or use uncertainty to support abstention and other reliability decisions~\cite{spiess2024calibration,zhou2024calibration,sharma2025assessing,pmlrjohnson23a,rathnasuriya2026answer}. Third, recent studies examine how uncertainty behaves when it is used to guide candidate selection, correction, or other generation-time decisions~\cite{shi2026code,rakasi2026uncertainty}. These studies show that uncertainty can provide useful information about correctness, but their evaluations mainly measure estimator performance or the effectiveness of an uncertainty-guided decision rule. However, important questions remain about the reliability of uncertainty as a proxy for execution correctness. Prior studies demonstrate that uncertainty can be informative, but they do not fully establish the extent to which it remains reliable when passing and failing generations receive similar uncertainty scores. It also remains unclear where this mismatch arises during generation, under which conditions it persists, and whether and to what extent common mitigation strategies can address it. Addressing these questions requires a systematic characterization of the reliability limits of model-derived uncertainty in code generation. 

To characterize the reliability limits of model-derived uncertainty, we first conduct a preliminary study of existing uncertainty measures (Section~\ref{sec:rq1}). Across 17 generation-stage metrics and six prefill-stage metrics, no measure provides stable failure ranking across models and benchmarks.  We then examine overconfidence through token confidence and predictive entropy at both program and token levels (Section~\ref{sec:rq2}). Failing programs often remain highly certain, including within their least-confident token regions. This misalignment affects selective generation where reducing coverage does not consistently increase selective accuracy and can decrease it when failures occupy high-certainty regions. Furthermore, instruction tuning can also increase certainty on failing generations without a corresponding improvement in correctness discrimination. These findings show that overconfidence is a distinct reliability problem where improvements in generation accuracy do not necessarily improve the model’s ability to identify which generations are likely correct.

Finally, we evaluate whether common mitigation techniques reduce overconfident failure (Section~\ref{sec:rq3_mitigations}). Post-hoc calibration substantially improves probability alignment, but calibrated scores provide little predictive skill beyond the benchmark pass rate. Repeated sampling improves recovery at small budgets, but its improvements diminish and a persistent set of tasks remains unsolved. Moreover, explicit abstention prompting likewise does not reliably induce deferral on incorrect generations. As an exploratory analysis, we find systematic pass–fail differences in latent-state geometry that are not reflected in output-token confidence. These results show that improving generation success or confidence calibration does not necessarily eliminate overconfident failure. 

This paper makes the following contributions:

\begin{itemize}

    \item We formulate the \textit{dilemma of overconfidence} in code LLMs as a barrier to practical uncertainty-aware code generation. 

    \item We show that existing uncertainty signals provide only partial and model-dependent evidence of execution failure across both prefill and generation stages. 

    \item We characterize overconfidence at program and token levels, and show that it can persist under selective generation and after instruction tuning. 

    \item We evaluate post-hoc calibration, repeated sampling, and explicit abstention, showing that none reliably resolves overconfident failure despite improving specific aspects of reliability. 

    \item We provide exploratory evidence that latent representations contain correctness-related structure not reflected in output uncertainty, motivating representation-aware reliability mechanisms. 

\end{itemize}

\section{ Problem Setup and Scope}
\label{problem}

\begin{figure*}[htbp]
    \centering

    \begin{subfigure}{0.4\textwidth}
        \centering
        \includegraphics[width=\linewidth]{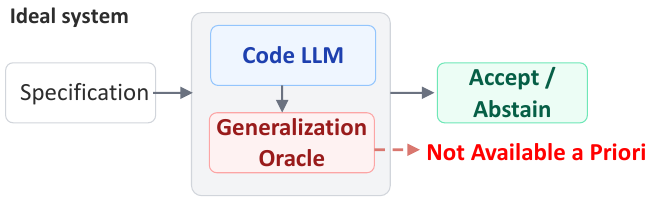}
        \caption{
        \textbf{Ideal oracle-based system.}
        A generalization oracle would determine whether generated code should be trusted, but such an oracle is not available a priori.
        }
        \label{fig:overview-ideal}
    \end{subfigure}
    \hspace{0.05\textwidth}
    \begin{subfigure}{0.4\textwidth}
        \centering
        \includegraphics[width=\linewidth]{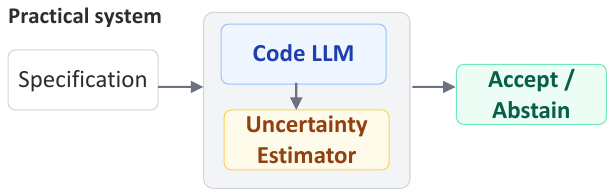}
        \caption{
        \textbf{Practical uncertainty-aware system.}
        The oracle is replaced by an uncertainty estimator that assigns a reliability score from model or generation behavior.
        }
        \label{fig:overview-practical}
    \end{subfigure}


    \begin{subfigure}{\textwidth}
        \centering
        \includegraphics[width=0.8\linewidth]{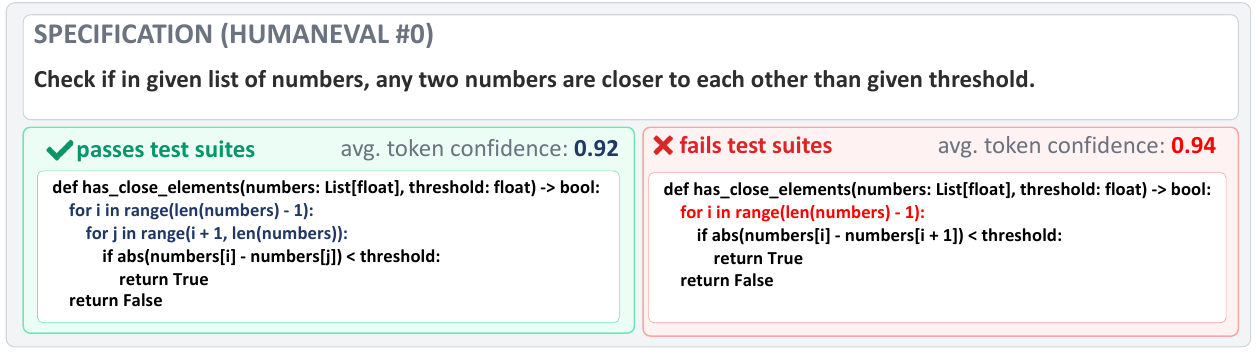}
        \caption{
        \textbf{Overconfident failure.}
        Two solutions for the same task receive similar mean token confidence, although one passes the test suite and the other fails by checking only adjacent elements on CodeLllama Model.
        }
        \label{fig:overview-example}
    \end{subfigure}

    \caption{
    Motivation for uncertainty-aware code generation.
    Since generalization oracle cannot be known a priori, practical systems use model-derived uncertainty as a proxy reliability signal. The example illustrates the failure mode studied in this paper: an incorrect program can receive confidence comparable to a passing program.
    }
    \label{fig:overview}
\end{figure*}

This section formalizes code generation as a selective decision problem and introduces the uncertainty-aware system view that motivates our study. We then define the sources of uncertainty signals and formulate overconfident failure as the central reliability problem.

\subsection{Code Generation as a Selective Decision Problem}


Given a prompt \(x\), a code-generation model \(M\) produces a candidate program \(y\). A task-specific test suite evaluates \(y\), yielding an execution outcome \(Z(x,y)\in\{\mathrm{pass},\mathrm{fail}\}\). We use this outcome as the operational correctness signal.


Standard code-generation evaluation measures whether a model can generate correct solutions, commonly through pass@\(k\). This measures \textbf{generation capability}, but does not indicate whether the model can distinguish its passing generations from its failing ones. We formulate the latter as a selective decision problem. Let \(S(x,y)\in\mathbb{R}\) denote a reliability score for candidate \(y\), where larger values indicate a greater estimated likelihood of passing. A selective system uses this score to apply a decision rule
\[
\pi(x,y,S)\in\{\mathrm{ACCEPT},\mathrm{ABSTAIN}\}.
\]


Thus, generation capability and reliability estimation are distinct. We study whether model-derived uncertainty scores provide reliable proxies for generation correctness.


\subsection{Uncertainty-Aware Code Generation}

Uncertainty-aware systems use model-derived evidence to decide whether an output should be trusted. In an ideal code-generation system (Figure~\ref{fig:overview-ideal}), a generalization oracle would determine whether a generated program generalizes to unseen tests, real inputs, or downstream use. Such an oracle is not available a priori, because functional correctness must be established through validation against specifications, tests, formal properties, or observed behavior.

Therefore, a practical uncertainty-aware system uses an estimated reliability signal as a proxy for unavailable correctness information. 
Figure~\ref{fig:overview-practical} illustrates this setting, where an uncertainty estimator assigns a score \(S(x,y)\) to generated code. The key assumption is that this score is informative about the execution outcome: higher-scoring candidates should be more likely to pass than lower-scoring candidates. Overconfident failure arises when this assumption breaks, such that a failing program receives a high reliability score and may therefore be accepted by the selective system.





\subsection{Sources of Uncertainty Signals}

We define an uncertainty-derived reliability signal as a measurable score \(S(x,y)\) attached to a generated candidate \(y\) for prompt \(x\). The score is intended to approximate the likelihood that \(y\) will pass the task's tests. 
We consider three sources of model-derived reliability signals. \textbf{White-box signals} use information from the model's generation process, including token probabilities, entropy, sequence likelihood, and hidden states. \textbf{Black-box signals} use variation across multiple sampled generations without requiring access to model internals. \textbf{Judge-based signals} use an external model to estimate the reliability of a generated program without executing it.




We also distinguish when the signal is measured. \emph{Prefill-stage} signals are computed before any code tokens are emitted, using only the prompt and the model's initial processing. They may capture prompt difficulty, ambiguity, or underspecification, but they cannot observe implementation-specific errors. \emph{Decoding-stage} signals are computed from generated tokens, token distributions, hidden states, or sampled candidates. These signals are closer to the final program, but they may still reflect local fluency or memorized coding patterns rather than functional correctness.



\subsection{The Problem: Overconfident Failure}

We define an \emph{overconfident failure} as a failing generation that receives a high reliability score. Formally, for prompt \(x\), candidate \(y\), execution outcome \(Z(x,y)\), and reliability score \(S(x,y)\), overconfident failure occurs when \(Z(x,y)=\mathrm{fail}\) while \(S(x,y)\) lies in a high-reliability region. Because score distributions vary across models and metrics, we do not assume a universal threshold. Instead, we examine whether failing generations receive reliability scores comparable to passing generations.

The underlying challenge is that model confidence and program correctness characterize different properties. Token-level confidence reflects the probability assigned to decoding decisions, whereas execution evaluates the semantics of the completed program. Therefore, a semantically incorrect implementation can receive high token-level confidence despite failing execution. Figure~\ref{fig:overview-example} illustrates this mismatch under the same specification: the passing implementation compares all element pairs, whereas the failing implementation checks only adjacent pairs. Despite this semantic difference, both generations receive comparable token confidence.

\section{Empirical Overview}
\label{sec:empirical_overview}


This section summarizes the study design, including the evaluated models, benchmarks, and research questions. Detailed experimental setups are described in the corresponding sections.


\subsection{Benchmarks}




We evaluate generated code on three execution-based benchmarks: \textsc{HumanEval+}, \textsc{MBPP+}, and \textsc{BigCodeBench}. HumanEval+~\cite{chen2021evaluating,liu2023your} and MBPP+~\cite{liu2023your,dong2025codescore} extend widely used function-level code-generation benchmarks with stronger test suites from EvalPlus~\cite{evalplus}. HumanEval+ contains 164 programming tasks specified by function signatures and natural-language descriptions. The MBPP+ evaluation set contains 378 tasks derived from introductory programming problems. These two benchmarks provide controlled function-level settings for evaluating execution correctness. BigCodeBench~\cite{zhuo2025bigcodebench} broadens the evaluation to programming tasks with richer contexts, more complex instructions, and greater use of library APIs. It contains 1,140 Python tasks spanning 723 function calls from 139 libraries across seven domains. We include BigCodeBench to examine whether overconfident failure persists beyond controlled function-level tasks.



\noindent\textbf{Models.}
We study four code-specialized, open-source instruction-tuned LLMs: \textit{CodeLlama-7B-Instruct}, \textit{DeepSeek-Coder-7B-Instruct}, \textit{Qwen2.5-Coder-7B-Instruct}, and \textit{OpenCoder-8B-Instruct}~\cite{roziere2023code,guo2024deepseek,zheng2024opencodeinterpreter}. These models have been used in prior code-generation evaluations and provide comparable open-source baselines in the 7-8B parameter range~\cite{liu2024exploring,liu2023your,wang2024testeval}.

We select these models based on four criteria. First, their open-source availability enables access to token-level statistics, prefill-stage signals, and internal representations. Second, restricting the models to 7-8B parameters reduces variation attributable to model scale. Third, all models are specialized and instruction-tuned for code generation. Fourth, the models originate from different development lineages, reducing the risk that the observed behavior is specific to one training pipeline. For the post-training analysis in Section~\ref{post_training}, we additionally compare base and instruction-tuned variants of \textit{OpenCoder} and \textit{DeepSeekCoder}. Table~\ref{Tab:pass1} reports the pass@1 accuracy of the primary models on each benchmark. All experiments were conducted on a server equipped with an Intel(R) Xeon(R) w5-2545 CPU and four NVIDIA RTX 6000 Ada GPUs.

\begin{table}[htb!]
    \centering
    \tabcolsep=1.4mm
    \caption{Pass@1 Accuracy for Each Benchmark and Models }
    \label{Tab:pass1}
    \renewcommand{\arraystretch}{}
    \resizebox{0.7\linewidth}{!}{
    \begin{tabular}{c c c c c}
    \hline
        Model & CodeLlama & DeepSeek-Coder & OpenCoder & QwenCoder \\ \hline
        HumanEval+ & 34.1 & 70.1 & 78.0 & 71.3 \\ \hline
        MBPP+ & 46.0 & 66.1 & 71.2 & 65.9 \\ \hline
        BigCodeBench &6.5&20.6&23.8 & 8.2  \\\hline
    \end{tabular}
    }
\end{table}

\subsection{Research Questions}
\label{rqs}


We investigate overconfident failure through the following research questions.



 

\noindent\textbf{RQ1:} To what extent do existing uncertainty signals reflect the likelihood of generation correctness?

\noindent\textbf{RQ2:} How does overconfident failure manifest, and under what conditions does it persist in code generation?

\noindent\textbf{RQ3:} To what extent do existing mitigation strategies reduce overconfident failure?

\noindent\textbf{RQ4:} Do latent representations contain correctness information not exposed by output uncertainty?

\section{Reliability of Standard Uncertainty Signals for Failure Detection}
\label{sec:rq1}


\begin{table}[t]
\centering
\caption{Uncertainty signals evaluated in the preliminary study.}
\label{tab:metric_families}

\renewcommand{\arraystretch}{1.0}
\small

\resizebox{0.6\columnwidth}{!}{%
\begin{tabular}{@{}p{2cm}p{4cm}p{6cm}@{}}
\toprule
\textbf{Family} & \textbf{Metrics} & \textbf{Signal captured} \\
\midrule

White-box token scores
& MSP, Perplexity, Entropy, PMI, CPMI, R-Div, F-R Dist~\cite{darrin2022rainproof,van2022mutual,takayama2019relevant,fomicheva2020unsupervised,vashurin2024benchmarking}
& Measure token likelihood, surprise, distributional concentration, and prompt dependence. \\

White-box multi-sample scores
& MCEnt, MSPS, RMI, Ens, MI\cite{kuhn2023semantic,vashurin2024benchmarking}
& Measure probability support and distributional disagreement across sampled candidates. \\

Black-box output similarity
& LexSim, EigV, DegGlobal~\cite{lin2023generating,fomicheva2020unsupervised}
& Measure diversity or disagreement among sampled programs without using model probabilities. \\

Judge-based scores
& Judge, M\_Judge~\cite{gu2024survey,vashurin2024benchmarking}
& Measure LLM-estimated pass likelihood from the prompt and generated solution. \\

\bottomrule
\end{tabular}%
}
\end{table}

This section addresses \textbf{RQ1} in Section~\ref{rqs} by examining to what extent standard uncertainty signals reliably indicate the likelihood that generated code will fail. This research question serves as an initial diagnostic analysis: if existing metrics consistently separate passing from failing programs, then overconfident failure would be less concerning for selective code generation. 

\textbf{Evaluation Setup.}
For each generated solution, we compute 17 generation-stage uncertainty scores. We also compute six prefill scores:  end-of-prompt next-token MSP, entropy, margin, energy, mean token entropy, and mean negative log-likelihood. Table~\ref{tab:metric_families} summarizes the evaluated metrics. For metrics that require multiple generations, we sample up to 10 candidates per prompt
at temperature 1.0.

We evaluate each score using AUC, with \texttt{fail} as the positive class, and orient the scores so that larger values indicate greater estimated failure risk. An AUC of 0.5 corresponds to random ranking. We focus on HumanEval+ and MBPP+ because both passing and failing generations are sufficiently represented across the studied models. We exclude BigCodeBench because its substantially lower pass rates leave few passing generations for several models, making pass-fail discrimination estimates less stable. Full metric definitions and implementation details are available on the project website~\cite{CodeCalibration}.

\textbf{Results.}
Table~\ref{tab:uncertainty_auc} reports the generation and prefill results. For generation-stage uncertainty, the best AUC in each model-benchmark pairs ranges from 0.598 to 0.696. No metric exceeds 0.70, and several remain close to random ranking. The strongest metric also changes across models and benchmarks where no metric family consistently provides the best separation.

Prefill uncertainty shows the same limitation. The highest AUC is 0.70 for CodeLlama on HumanEval+, whereas most model--benchmark settings fall between 0.52 and 0.62. Several scores are close to random ranking, particularly on MBPP+, and no prefill metric is consistently strongest across models.
\begin{table}[t]
\centering
\caption{AUC ($\uparrow$) results of prefill-stage and generative uncertainty metrics.}
\label{tab:uncertainty_auc}

\renewcommand{\arraystretch}{}
\small

\resizebox{0.8\columnwidth}{!}{%
\begin{tabular}{lcccccccc}
\toprule
\multirow{2}{*}{Metric}
& \multicolumn{4}{c}{HumanEval+}
& \multicolumn{4}{c}{MBPP+} \\
\cmidrule(lr){2-5}\cmidrule(lr){6-9}
& CodeLlama & DeepSeek & OpenCoder & QwenCoder
& CodeLlama & DeepSeek & OpenCoder & QwenCoder \\
\midrule

\multicolumn{9}{l}{\textbf{Prefill uncertainty}} \\
\midrule

MSP
& 0.690 & \textbf{0.590} & 0.530 & 0.530
& 0.460 & 0.490 & 0.480 & 0.610 \\

Entropy
& \textbf{0.700} & \textbf{0.590} & 0.530 & 0.540
& 0.470 & \textbf{0.600} & 0.500 & 0.540 \\

Margin
& 0.690 & \textbf{0.590} & 0.520 & 0.520
& 0.470 & 0.450 & 0.480 & \textbf{0.620} \\

Energy
& 0.590 & 0.420 & 0.540 & \textbf{0.560}
& 0.530 & 0.540 & 0.510 & 0.520 \\

MeanTokenEntropy
& 0.600 & 0.580 & 0.540 & 0.430
& 0.560 & 0.570 & 0.540 & 0.540 \\

NLL
& 0.610 & \textbf{0.590} & \textbf{0.580} & 0.440
& \textbf{0.590} & \textbf{0.600} & \textbf{0.570} & 0.550 \\

\midrule
\multicolumn{9}{l}{\textbf{Generative uncertainty}} \\
\midrule

MSP
& \textbf{0.670} & 0.623 & 0.630 & 0.617
& 0.568 & 0.561 & 0.517 & 0.622 \\

Perplexity
& \textbf{0.670} & 0.623 & 0.630 & 0.617
& 0.568 & 0.561 & 0.517 & 0.622 \\

Entropy
& 0.630 & 0.615 & 0.639 & 0.635
& 0.570 & 0.567 & 0.524 & \textbf{0.629} \\

PMI
& 0.521 & 0.510 & 0.501 & 0.557
& 0.589 & 0.572 & 0.546 & 0.569 \\

CPMI
& 0.506 & 0.589 & 0.503 & \textbf{0.681}
& 0.578 & 0.509 & 0.551 & 0.546 \\

R-Div
& 0.662 & 0.619 & \textbf{0.640} & 0.625
& 0.573 & 0.567 & 0.519 & 0.623 \\

F-R Dist
& 0.568 & 0.614 & 0.627 & 0.640
& 0.513 & 0.543 & 0.533 & \textbf{0.629} \\

MCEnt
& 0.630 & 0.553 & 0.622 & 0.594
& 0.517 & 0.598 & 0.588 & 0.529 \\

MSPS
& 0.623 & 0.554 & 0.622 & 0.589
& 0.507 & 0.599 & 0.589 & 0.533 \\

RMI
& 0.508 & 0.542 & 0.617 & 0.599
& 0.582 & 0.578 & 0.638 & 0.511 \\

Ens
& 0.579 & 0.628 & 0.624 & 0.615
& \textbf{0.598} & \textbf{0.696} & \textbf{0.660} & 0.565 \\

MI
& 0.577 & 0.629 & 0.618 & 0.634
& 0.596 & 0.691 & 0.651 & 0.600 \\

LexSim
& 0.596 & 0.629 & 0.611 & 0.536
& 0.568 & 0.589 & 0.615 & 0.515 \\

EigV
& 0.597 & \textbf{0.632} & 0.609 & 0.532
& 0.563 & 0.585 & 0.609 & 0.516 \\

DegGlobal
& 0.576 & 0.568 & 0.521 & 0.574
& 0.578 & 0.561 & 0.556 & 0.514 \\

Judge
& 0.530 & 0.514 & 0.601 & 0.507
& 0.542 & 0.538 & 0.631 & 0.522 \\

M\_Judge
& 0.530 & 0.514 & 0.601 & 0.506
& 0.542 & 0.538 & 0.631 & 0.522 \\

\bottomrule
\end{tabular}%
}
\end{table}

\noindent\textbf{Finding.}
The limited failure discrimination is not specific to one uncertainty formulation or one stage of inference. Across 17 generation-stage signals and six prefill signals, no uncertainty measure provides stable failure ranking
across models and benchmarks.

\noindent\textbf{Implication.}
The results do not support a model- and task-independent mapping from standard uncertainty scores to code correctness. Furthermore, an uncertainty signal
that is useful in one setting cannot be assumed to provide the same failure evidence in another.

\noindent\textbf{SE Takeaway.} Code-generation tools should not reuse uncertainty thresholds across models or workloads. Acceptance and validation policies should be re-evaluated for each deployment setting because the same uncertainty signal can provide different failure discrimination across settings.

\section{Characterizing Overconfidence During Code Generation}
\label{sec:rq2}


In this section, we answer \textbf{RQ2} by characterizing how overconfidence  manifests across the code-generation process. We examine whether failing programs remain highly confident overall, whether local token behavior exposes regions of uncertainty, and whether predictive entropy provides additional evidence of failure. We further study the effect of these uncertainty signals on selective accuracy and examine whether overconfidence persists after instruction tuning.







\subsection{Program-Level Confidence Remains High for Failing Generations}

\begin{table*}[t]
\centering
\caption{Statistics on average and bottom-10\% token confidence for passing and failing generated programs.}
\label{tab:confidence_full_stats}

\renewcommand{\arraystretch}{}
\scriptsize

\resizebox{\linewidth}{!}{%
\begin{tabular}{llrr@{\hspace{8pt}}rrrrrr@{\hspace{10pt}}rrrrrr}
\toprule
\multirow{3}{*}{Task} & \multirow{3}{*}{Model}
& \multicolumn{2}{c}{Sample Size}
& \multicolumn{6}{c}{Average Confidence}
& \multicolumn{6}{c}{Bottom-10\% Confidence} \\
\cmidrule(lr){3-4} \cmidrule(lr){5-10} \cmidrule(lr){11-16}
&
& \multirow{2}{*}{Correct $n$} & \multirow{2}{*}{Incorrect $n$}
& \multicolumn{2}{c}{Correct}
& \multicolumn{2}{c}{Incorrect}
& \multirow{2}{*}{t-value}
& \multirow{2}{*}{p-value}
& \multicolumn{2}{c}{Correct}
& \multicolumn{2}{c}{Incorrect}
& \multirow{2}{*}{t-value}
& \multirow{2}{*}{p-value} \\
\cmidrule(lr){5-6} \cmidrule(lr){7-8}
\cmidrule(lr){11-12} \cmidrule(lr){13-14}
&
& &
& Mean $\pm$ Std. & Median
& Mean $\pm$ Std. & Median
& &
& Mean $\pm$ Std. & Median
& Mean $\pm$ Std. & Median
& & \\
\midrule

\multirow{4}{*}{HumanEval+}
& CodeLlama
& 56 & 108
& $0.962 \pm 0.012$ & 0.965
& $0.956 \pm 0.014$ & 0.957
& 3.194 & 0.002
& $0.703 \pm 0.076$ & 0.721
& $0.651 \pm 0.083$ & 0.654
& 4.054 & 0.000 \\

& DeepSeek
& 115 & 49
& $0.951 \pm 0.021$ & 0.957
& $0.943 \pm 0.021$ & 0.946
& 2.233 & 0.028
& $0.642 \pm 0.090$ & 0.658
& $0.607 \pm 0.075$ & 0.610
& 2.548 & 0.012 \\

& OpenCoder
& 128 & 36
& $0.956 \pm 0.024$ & 0.960
& $0.948 \pm 0.023$ & 0.953
& 1.941 & 0.057
& $0.675 \pm 0.119$ & 0.677
& $0.631 \pm 0.090$ & 0.635
& 2.403 & 0.019 \\

& QwenCoder
& 117 & 47
& $0.947 \pm 0.025$ & 0.953
& $0.935 \pm 0.036$ & 0.943
& 2.047 & 0.045
& $0.619 \pm 0.101$ & 0.640
& $0.559 \pm 0.116$ & 0.570
& 3.106 & 0.003 \\

\midrule

\multirow{4}{*}{MBPP+}
& CodeLlama
& 174 & 204
& $0.925 \pm 0.025$ & 0.930
& $0.919 \pm 0.026$ & 0.921
& 2.312 & 0.021
& $0.542 \pm 0.081$ & 0.543
& $0.525 \pm 0.083$ & 0.522
& 1.991 & 0.047 \\

& DeepSeek
& 250 & 128
& $0.944 \pm 0.023$ & 0.947
& $0.938 \pm 0.026$ & 0.940
& 2.301 & 0.022
& $0.609 \pm 0.096$ & 0.606
& $0.589 \pm 0.102$ & 0.586
& 1.843 & 0.067 \\

& OpenCoder
& 269 & 109
& $0.924 \pm 0.024$ & 0.927
& $0.922 \pm 0.028$ & 0.924
& 0.515 & 0.607
& $0.550 \pm 0.080$ & 0.549
& $0.542 \pm 0.095$ & 0.540
& 0.797 & 0.426 \\

& QwenCoder
& 249 & 129
& $0.901 \pm 0.035$ & 0.902
& $0.881 \pm 0.049$ & 0.886
& 4.100 & 0.000
& $0.480 \pm 0.086$ & 0.473
& $0.439 \pm 0.117$ & 0.440
& 3.485 & 0.001 \\

\midrule

\multirow{4}{*}{BigCodeBench}
& CodeLlama
& 75 & 1065
& $0.943 \pm 0.020$ & 0.946
& $0.956 \pm 0.039$ & 0.959
& -5.171 & 0.000
& $0.602 \pm 0.094$ & 0.585
& $0.730 \pm 0.204$ & 0.668
& -10.249 & 0.000 \\

& DeepSeek
& 235 & 905
& $0.913 \pm 0.030$ & 0.917
& $0.909 \pm 0.031$ & 0.914
& 1.587 & 0.113
& $0.511 \pm 0.080$ & 0.507
& $0.498 \pm 0.079$ & 0.501
& 2.289 & 0.023 \\

& OpenCoder
& 272 & 868
& $0.936 \pm 0.025$ & 0.938
& $0.934 \pm 0.027$ & 0.937
& 0.908 & 0.365
& $0.580 \pm 0.099$ & 0.575
& $0.577 \pm 0.095$ & 0.572
& 0.428 & 0.669 \\

& QwenCoder
& 94 & 1046
& $0.897 \pm 0.041$ & 0.897
& $0.920 \pm 0.068$ & 0.931
& -4.746 & 0.000
& $0.455 \pm 0.122$ & 0.434
& $0.574 \pm 0.223$ & 0.516
& -8.366 & 0.000 \\

\bottomrule
\end{tabular}%
}

\vspace{2pt}
\begin{minipage}{0.98\textwidth}
\footnotesize
 $p=0.000$ indicates $p<0.001$ due to rounding.
\end{minipage}

\end{table*}

\textbf{Evaluation Setup.}
We next measure confidence over the generated code. For each solution, we compute average token confidence as the mean probability assigned to emitted tokens. To test whether failures are exposed by localized uncertainty, we also compute bottom-10\% confidence, defined as the mean confidence of the least confident 10\% of generated tokens. We compare passing and failing generations using mean, standard deviation, median, and Welch's two-sample (t)-test.


\textbf{Results.}
Table~\ref{tab:confidence_full_stats} shows that both correct and incorrect programs receive high average token confidence. On HumanEval+ and MBPP+, average confidence is typically above 0.90 for both groups. Although correct solutions often have slightly higher confidence, the absolute gaps are small. For example, on HumanEval+, the correct--incorrect average-confidence gap ranges from 0.007 to 0.012 across models. On MBPP+, the gap is even smaller for several models, including OpenCoder, where correct and incorrect generations have nearly identical average confidence.

The pattern becomes more striking on BigCodeBench. For CodeLlama and QwenCoder, incorrect generations have \emph{higher} average confidence than correct generations. CodeLlama assigns mean confidence 0.956 to incorrect generations and 0.943 to correct generations, while QwenCoder assigns 0.920 to incorrect generations and 0.897 to correct generations. These negative $t$-statistics, which are statistically significant under standard two-sample $t$-tests, indicate that the observed differences are unlikely to be due to random variation; rather, they reveal a systematic confidence--failure mismatch, where the models are most confident precisely when they are wrong.


\noindent\textbf{Finding.}
Generation-time confidence is high for both passing and failing programs. On controlled benchmarks, confidence gaps are small; on BigCodeBench, some models are more confident on failures than on correct solutions.

\subsection{Local Token Confidence Provides Limited Failure Evidence}



Program-level confidence may hide isolated generation steps at which the model is uncertain. Therefore, we examine the least-confident tokens (i.e., the bottom 10\% by confidence) within each generated program to determine whether failures contain localized uncertainty that is not reflected in the overall confidence score.

Table~\ref{tab:confidence_full_stats} shows that bottom-10\% confidence is lower than average confidence, as expected, but it still does not reliably separate correct and incorrect programs. On HumanEval+, correct solutions have higher bottom-10\% confidence than incorrect solutions, but the distributions remain close. On MBPP+, the difference is weak for several models and statistically insignificant for OpenCoder. On BigCodeBench, the trend reverses for CodeLlama and QwenCoder: incorrect generations have substantially higher bottom-10\% confidence than correct generations.Thus, local token confidence provides evidence in some settings, but many incorrect programs remain locally confident even at their least confident tokens.

\noindent\textbf{Finding.}
Local token confidence is insufficient: incorrect programs can remain confident even in their least confident token regions.
\begin{figure}[t]
\centering

\begin{subfigure}[t]{0.28\columnwidth}
    \centering
    \includegraphics[width=\linewidth]{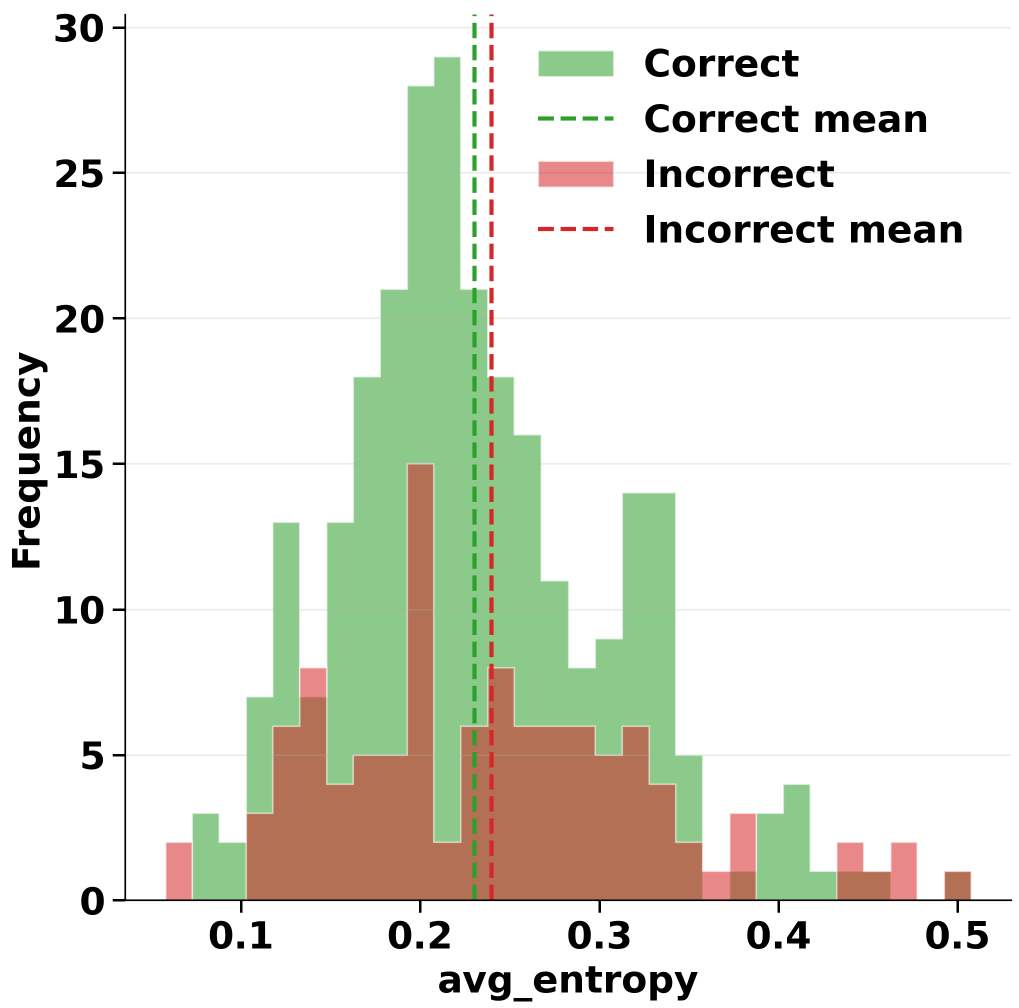}
    \caption{Average token entropy}
    \label{fig:opencoder_avg_entropy}
\end{subfigure}
\begin{subfigure}[t]{0.28\columnwidth}
    \centering
    \includegraphics[width=\linewidth]{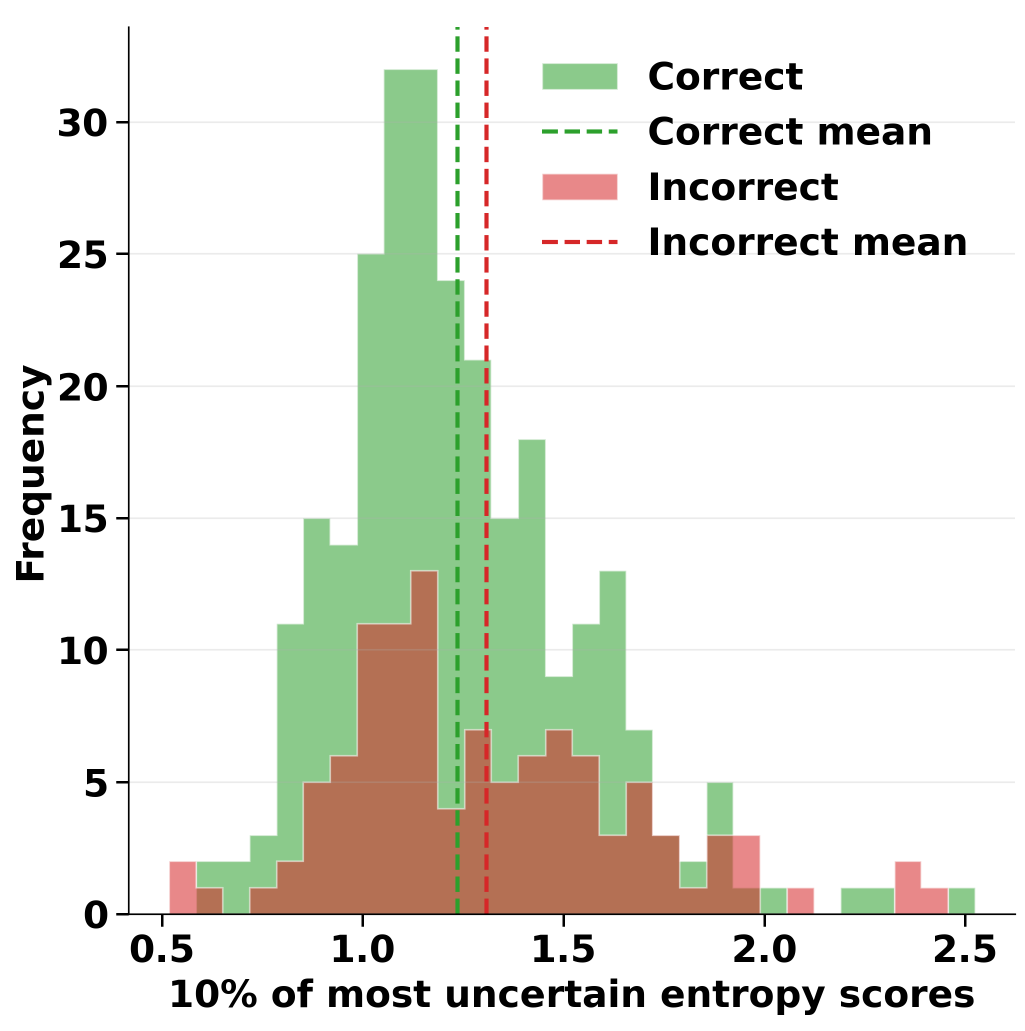}
    \caption{Top-10\% token entropy}
    \label{fig:opencoder_top10_entropy}
\end{subfigure}

\caption{Entropy distributions for OpenCoder on MBPP+.}
\label{fig:opencoder_entropy_mbpp}
\end{figure}

\subsection{Predictive Entropy Provides Limited Pass-Fail Separation}



We further examine predictive entropy as a complementary measure of generation-time uncertainty. At each generation step, predictive entropy is computed from the model's probability distribution over all possible next tokens. Entropy is low when probability mass is concentrated on a small set of candidates and high when it is distributed more broadly across alternatives. Unlike token confidence, which considers only the probability assigned to the emitted token, predictive entropy accounts for the full next-token distribution. We test whether this broader uncertainty signal better distinguishes failing from passing generations. If predictive entropy reflects functional correctness, failing generations should exhibit systematically higher entropy than passing generations.


Figure~\ref{fig:opencoder_entropy_mbpp} presents representative entropy distributions for OpenCoder on MBPP+. Both average entropy and top-10\% entropy show substantial overlap between correct and incorrect generations. Although the mean entropy for incorrect outputs is marginally higher, the distributions are not clearly separable. This observation aligns with the confidence analysis such that entropy captures some uncertainty, but not enough to reliably identify failing programs.

\noindent\textbf{Finding.}
Entropy provides the same qualitative picture as confidence. Failing programs often remain in low-uncertainty regions, and local high-entropy scores do not consistently isolate failures.

\begin{figure*}[t]
    \centering

    \begin{subfigure}[t]{0.24\textwidth}
        \centering
        \includegraphics[width=\linewidth]{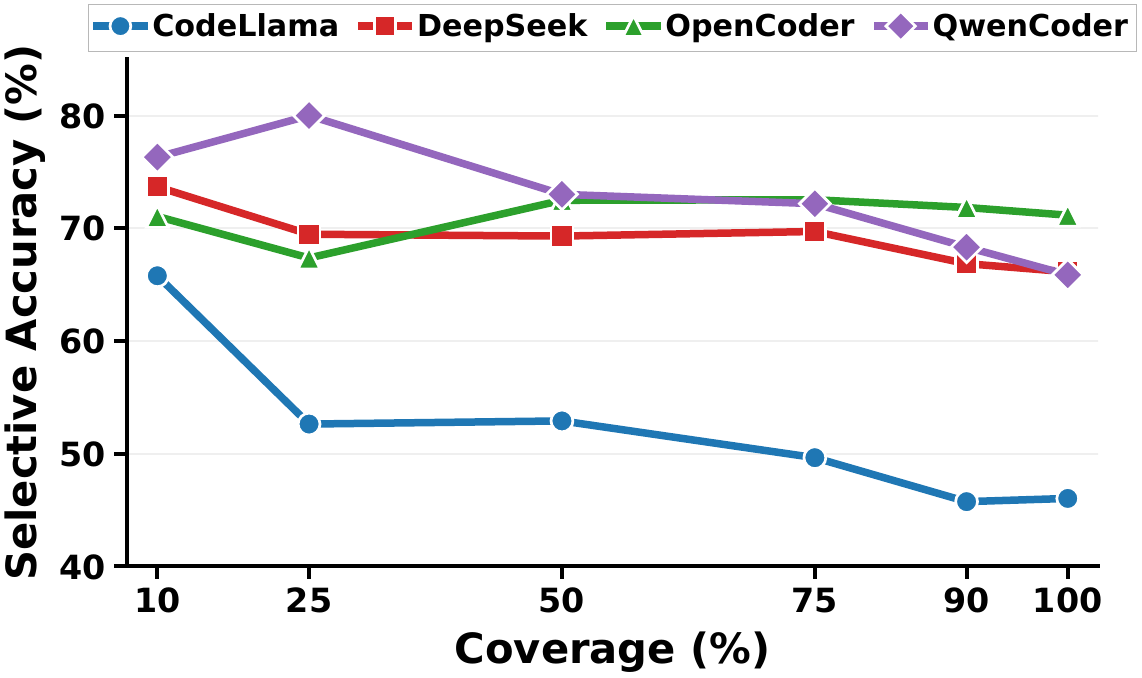}
        \caption{MBPP+ with avg. confidence}
        \label{fig:selective_mbpp_conf}
    \end{subfigure}\hfill
    \begin{subfigure}[t]{0.24\textwidth}
        \centering
        \includegraphics[width=\linewidth]{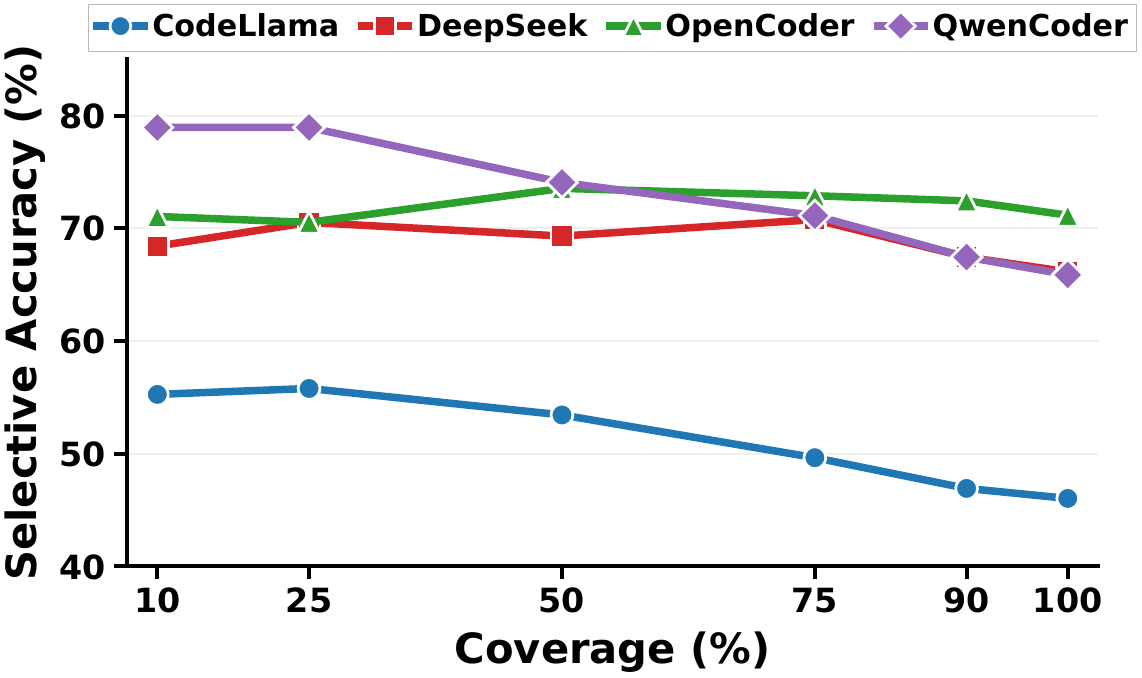}
        \caption{MBPP+ with avg. entropy}
        \label{fig:selective_mbpp_ent}
    \end{subfigure}\hfill
    \begin{subfigure}[t]{0.24\textwidth}
        \centering
        \includegraphics[width=\linewidth]{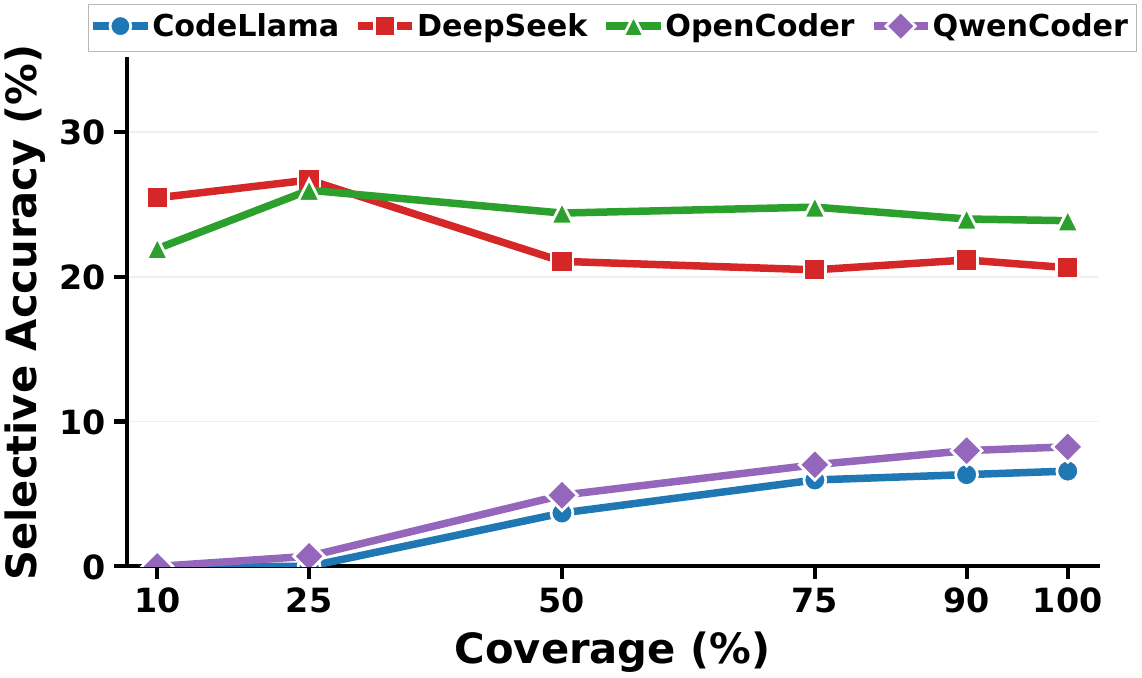}
        \caption{BigCodeBench with avg. confidence}
        \label{fig:selective_bcb_conf}
    \end{subfigure}\hfill
    \begin{subfigure}[t]{0.24\textwidth}
        \centering
        \includegraphics[width=\linewidth]{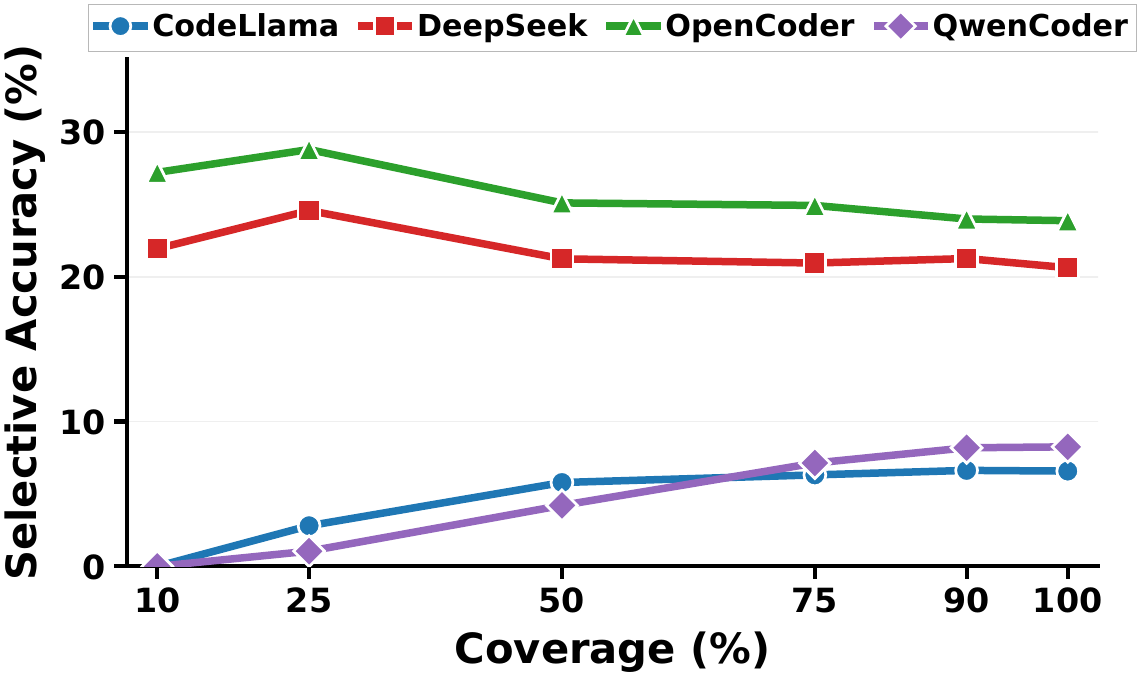}
        \caption{BigCodeBench with avg. entropy}
        \label{fig:selective_bcb_ent}
    \end{subfigure}

    \caption{Selective accuracy under uncertainty-based ranking. At each coverage level, we retain the most certain generations according to average confidence or average entropy and report their execution accuracy. Lower coverage corresponds to accepting a smaller subset of generations.}
    \label{fig:selective_accuracy}
\end{figure*}

\subsection{Uncertainty-Based Selective Generation}

We next evaluate whether uncertainty-based ranking can identify subsets of generated programs with higher execution accuracy. This analysis investigates whether overconfident failures persist among the generations assigned the highest certainty and therefore remain eligible for acceptance.

\textbf{Evaluation Setup.}
For each generated program, we use average confidence and average entropy as selection scores. We rank generations from most to least certain and retain the top $c\%$ at coverage $c \in \{10,25,50,75,90,100\}$. For example, 10\% coverage accepts only the 10\% most certain generations and rejects the remaining 90\%. We measure execution accuracy among the accepted programs.

Therefore, selective generation trades coverage for accuracy. If uncertainty reflects functional correctness, removing uncertain generations should increase the accuracy of the accepted subset relative to the model's original accuracy. If high-certainty failures remain, reducing coverage may provide little improvement or even reduce accuracy.

\textbf{Results.}
Figure~\ref{fig:selective_accuracy} shows different selection behavior across models and benchmarks. On MBPP+, uncertainty identifies more accurate subsets for CodeLlama, DeepSeekCoder, and QwenCoder. For CodeLlama, confidence increases accuracy from 46.0\% at 100\% coverage to 65.79\% at 10\% coverage. QwenCoder increases from 65.9\% to 80.00\% at 25\% coverage. Entropy produces a similar behavior for QwenCoder, reaching 78.95\% at both 10\% and 25\% coverage. DeepSeekCoder shows only a marginal increase. However, for OpenCoder its original accuracy is 71.2\%, yet confidence yields 71.05\% at 10\% coverage and 67.37\% at
25\%. Likewise, entropy provides similarly marginal improvement.

BigCodeBench shows that this behavior does not persist across benchmarks. DeepSeekCoder and OpenCoder achieve modest accuracy improvements after uncertainty-based filtering. In contrast, the 10\% most certain generations from CodeLlama and QwenCoder contain no passing solutions under either confidence or entropy where their original accuracies are 6.5\% and 8.2\%, respectively. For CodeLlama, no generation in the top 25\% by confidence passes either. In these cases, uncertainty-based filtering reduces accuracy rather than improving it.

The comparison further separates uncertainty-ranking quality from generation accuracy. OpenCoder has the highest original accuracy on MBPP+ but receives little accuracy improvement from selective generation. Conversely, models with lower original accuracy can obtain substantially more accurate retained subsets.

\noindent\textbf{Finding.}
Overconfident failures can occupy the high-certainty region of the uncertainty ranking. As a result, decreasing coverage does not consistently improve execution accuracy and can reduce it when uncertainty is misaligned with functional correctness. This misalignment is model- and benchmark-dependent and does not follow the model's original generation accuracy.

\begin{figure*}[t]
    \centering

    \begin{subfigure}[t]{0.24\textwidth}
        \centering
        \includegraphics[width=\linewidth]{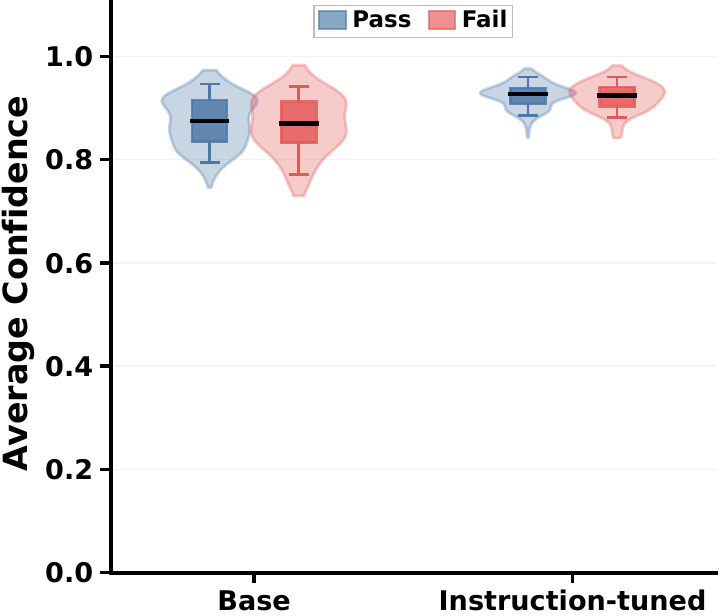}
        \caption{OpenCoder with avg. confidence}
        \label{fig:posttraining_opencoder_conf}
    \end{subfigure}\hfill
    \begin{subfigure}[t]{0.24\textwidth}
        \centering
        \includegraphics[width=\linewidth]{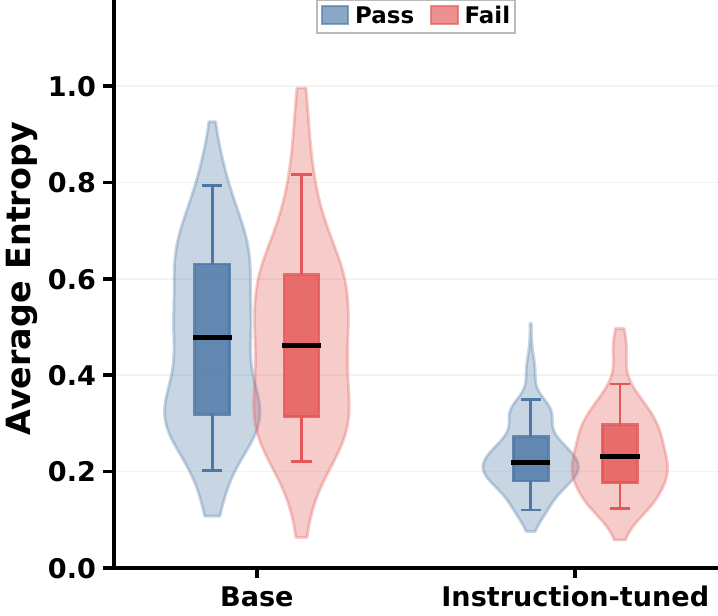}
        \caption{OpenCoder with avg. entropy}
        \label{fig:posttraining_opencoder_ent}
    \end{subfigure}\hfill
    \begin{subfigure}[t]{0.24\textwidth}
        \centering
        \includegraphics[width=\linewidth]{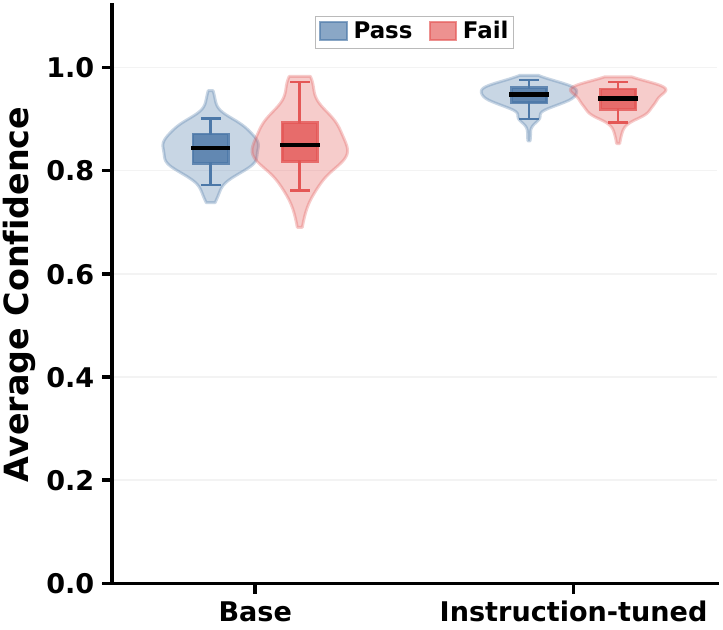}
        \caption{DeepSeekCoder with avg. confidence}
        \label{fig:posttraining_deepseek_conf}
    \end{subfigure}\hfill
    \begin{subfigure}[t]{0.24\textwidth}
        \centering
        \includegraphics[width=\linewidth]{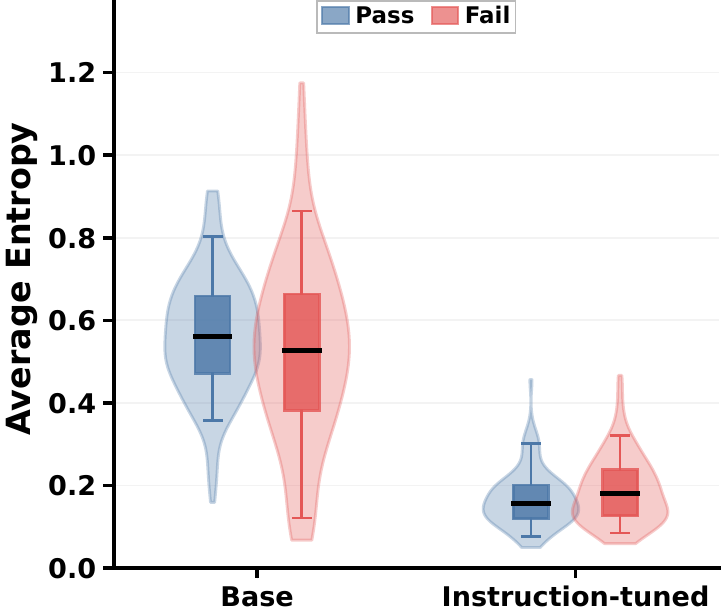}
        \caption{DeepSeekCoder with avg. entropy}
        \label{fig:posttraining_deepseek_ent}
    \end{subfigure}

    \caption{Uncertainty before and after instruction tuning on MBPP+. Each plot compares the distributions of passing and failing generations for the corresponding base and instruction-tuned model.}
    \label{fig:posttraining_uncertainty}
\end{figure*}

\subsection{Instruction Tuning Increases Certainty in Failing Generations}
\label{post_training}

We next examine how instruction tuning changes overconfidence relative to the corresponding base model. Specifically, we distinguish changes in the certainty assigned to failing generations from changes in how well uncertainty separates passing from failing programs. This distinction allows us to determine whether post-training improves uncertainty-based self-assessment or merely changes the magnitude of model certainty.


\textbf{Evaluation Setup.}
We compare base and instruction-tuned variants of OpenCoder and DeepSeekCoder on MBPP+. We evaluate average confidence and average entropy for passing and failing generations and quantify their separation using AUC. To assess whether the change in discrimination after instruction tuning is stable across tasks, we compute a paired task-level bootstrap 95\% confidence interval for $\Delta$AUC, defined as the instruction-tuned AUC minus the base-model AUC.

\textbf{Results.}
Figure~\ref{fig:posttraining_uncertainty} shows how instruction tuning changes the confidence and entropy distributions of passing and failing generations for
OpenCoder and DeepSeekCoder. Instruction tuning increases certainty on failing generations for both model families. For OpenCoder, the average confidence of failing programs increases from 0.8685 to 0.9224, while their average entropy decreases from 0.4754 to 0.2401. Passing generations exhibit a comparable shift. As a result, the additional certainty does not increase pass--fail separation. Confidence AUC changes from 0.5230 to 0.5094, with $\Delta$AUC $=-0.0136$ and a 95\% confidence interval of $[-0.0911, 0.0652]$. Entropy AUC changes from 0.4915 to 0.5237, with $\Delta$AUC $=+0.0323$ and an interval of $[-0.0473, 0.1135]$. These intervals provide no evidence of a systematic change in discrimination.

For DeepSeekCoder, instruction tuning also increases certainty on failing programs, but the uncertainty- correctness ordering changes. In the base model, confidence and entropy AUCs are 0.4391 and 0.4415, indicating that greater estimated certainty is associated with failing generations. After instruction tuning, the AUCs increase to 0.5654 and 0.5672. The corresponding $\Delta$AUC values are $+0.1263$ for confidence and $+0.1257$ for entropy, with 95\% confidence intervals of $[0.0447, 0.2089]$ and $[0.0442, 0.2081]$, respectively. Therefore, instruction tuning  changes the ordering from negatively aligned to positively aligned with correctness. However, the resulting AUCs remain below 0.57, indicating substantial pass-fail overlap.

\noindent\textbf{Finding.}
Instruction tuning can increase overconfidence relative to the base model by increasing certainty on failing generations. However, this change in certainty does not determine whether uncertainty becomes more or less informative about functional correctness. For example, DeepSeekCoder becomes more certain on failing generations while its pass-fail discrimination improves. Thus, post-training can increase confidence in incorrect code independently of the reliability of uncertainty as a correctness signal.

\subsection{Implications of Overconfidence Characterization}

Our characterization separates three properties of reliability-aware code generation: functional accuracy, uncertainty-based correctness discrimination, and its stability after post-training. Pass/fail evaluation measures only the first and the latter two determine whether model uncertainty remains informative
for reliability decisions.


\textbf{Reliability evaluation should measure correctness discrimination separately from task accuracy.}
A model can achieve higher Pass@1 without becoming better at identifying which of its generations are correct. Therefore, evaluation should report whether uncertainty separates passing from failing programs in addition to reporting functional accuracy. Coverage-accuracy analysis provides a corresponding view by showing whether uncertainty ranking yields more accurate accepted subsets.

\textbf{Post-training objectives for correctness do not necessarily improve self-assessment.}
Our base-instruction comparison shows that model certainty and correctness discrimination can change independently after instruction tuning. Therefore, improving generation quality does not impose a corresponding constraint on confidence assigned to remaining failures. Reliability oriented post-training should treat generation correctness and uncertainty-based self-assessment as
separate optimization targets.

\textbf{Uncertainty behavior should be re-evaluated after model updates.}
Post-training can alter both the magnitude and the ordering of uncertainty scores. An uncertainty-based selection mechanism validated for one model version may no longer exhibit the same correctness discrimination after the model is updated. Model evolution should include regression evaluation of both functional accuracy and uncertainty-correctness alignment.

\noindent\textbf{SE Takeaway.}
Overconfidence is not reduced automatically by improving code-generation accuracy. Post-training can increase functional performance while high-certainty failures persist, making overconfidence a distinct reliability property that must be assessed independently.

\section{Limits of Mitigating Overconfident Failure}
\label{sec:rq3_mitigations}

This section answers \textbf{RQ3} on the effectiveness of  common mitigation techniques for reducing overconfident failure.  We evaluate three practical responses: post-hoc calibration, repeated sampling, and explicit abstention prompting. These strategies address different parts of the reliability problem where calibration rescales scores, sampling increases the chance of finding a correct candidate, and abstention prompting gives the model an explicit option to defer.

\subsection{Calibration Improves Probability Alignment but Provides Limited Predictive Skill}
\label{sec:calibration}

\textbf{Evaluation Setup.}
We examine to what extent post-hoc calibration mitigates overconfidence by mapping uncertainty scores to estimated probabilities of execution correctness. For confidence, we use average token confidence and bottom-10\% token confidence. For entropy, we use average token entropy and top-10\% token entropy. We calibrate each score using \textit{Platt scaling} (PS)~\cite{spiess2024calibration} which fits a logistic mapping from each original uncertainty score to an estimated probability of correctness and \textit{isotonic regression}~\cite{guo2017calibration} which fits a monotonic non-parametric mapping from uncertainty scores to empirical correctness. 

We estimate calibrated probabilities using five-fold stratified cross-validation to avoid evaluating a calibrator on the generations used to fit it. Each calibrator is trained on 80\% of the generations and applied to the held-out 20\%. For the uncalibrated baseline, we use confidence measures directly as confidence proxies and transform entropy $H$ to $e^{-H}$.

We use Brier score (BS) to measure the error between predicted pass probability and execution outcome~\cite{vashurin2024benchmarking,fomicheva2020unsupervised,fadeeva2024fact,takayama2019relevant,van2022mutual,darrin2022rainproof,duan2023shifting,kuhn2023semantic,lin2023generating,spiess2024calibration,devic2025calibration}. Lower BS indicates better probability alignment. Because a constant predictor equal to the benchmark pass rate can also achieve a low BS, we additionally report Brier Skill Score (BSS)~\cite{spiess2024calibration}. For pass rate $p_r$, the reference score is $B_{\mathrm{ref}}=p_r(1-p_r)$, and the skill score is $BSS=(B_{\mathrm{ref}}-BS)/B_{\mathrm{ref}}$. Positive BSS indicates improvement over the base-rate predictor, whereas zero indicates equivalent performance.

\textbf{Results.}
Table~\ref{tab:calibration_bs_bss} reports the calibration results on MBPP+ and BigCodeBench. Across both benchmarks, calibration substantially reduces Brier score, but the resulting Brier Skill Scores remain minimal. On MBPP+, calibration substantially reduces BS, but the calibrated probabilities provide little skill beyond the benchmark pass rate. For CodeLlama, PS reduces the BS of average token confidence from 0.459 to 0.248, while BSS is approximately 0.001. DeepSeekCoder shows a similar pattern where BS decreases from 0.300 to 0.224, but BSS remains approximately 0.001. For OpenCoder, BS decreases from 0.250 to 0.205 while BSS remains near zero. QwenCoder obtains higher positive skill in some settings, although its largest BSS is only 0.041 for bottom-10\% token confidence after isotonic regression.

On BigCodeBench, calibration produces large reductions in BS while BSS remains near zero across models. For CodeLlama, PS reduces the BS of average token confidence from 0.856 to 0.061, but the corresponding BSS is approximately 0.001. Therefore, the calibrated BS is nearly equivalent to the error of the base-rate predictor. The remaining models exhibit the same pattern, with calibrated BSS values generally close to zero and the largest values reaching only about 0.01--0.02.

The BS and BSS results show that calibration primarily corrects the scale of the uncertainty scores rather than providing substantial predictive information about individual generations. This interpretation is consistent with our selective-accuracy analysis, where passing and failing generations remain overlapping in the uncertainty space. Post-hoc calibration can remap these scores to better-aligned probabilities, but it cannot remove the underlying overlap between correct and incorrect generations.

\textbf{Finding.}
Post-hoc calibration corrects the magnitude of overconfidence but does not resolve overconfident failures. Although calibration substantially improves probability alignment, the calibrated scores provide little skill beyond the benchmark pass rate and remain unable to reliably distinguish passing from failing generations.




\begin{table*}[t]
\centering
\caption{Calibration results measured by Brier Score (BS; $\downarrow$) and
Brier Skill Score (BSS; $\uparrow$). Each entry reports BS / BSS.
For each model and metric, the lowest BS and highest BSS across
Before, PS, and Iso are shown in bold.}
\label{tab:calibration_bs_bss}

\setlength{\tabcolsep}{3.5pt}
\renewcommand{\arraystretch}{}
\small

\resizebox{\textwidth}{!}{%
\begin{tabular}{llcccccccc}
\toprule
& &
\multicolumn{4}{c}{MBPP+} &
\multicolumn{4}{c}{BigCodeBench} \\
\cmidrule(lr){3-6}
\cmidrule(lr){7-10}

Model & Method
& Avg. Conf.
& Avg. Ent.
& Bottom-10\% Conf.
& Top-10\% Ent.
& Avg. Conf.
& Avg. Ent.
& Bottom-10\% Conf.
& Top-10\% Ent. \\

\midrule


\multirow{3}{*}{CodeLlama}
& Before
& 0.459 / -0.849
& 0.345 / -0.387
& 0.252 / -0.015
& 0.284 / -0.142
& 0.856 / -12.926
& 0.699 / -10.379
& 0.548 / -7.922
& 0.343 / -4.578 \\

& PS
& \textbf{0.248} / \textbf{+0.001}
& \textbf{0.247} / \textbf{+0.005}
& \textbf{0.248} / \textbf{+0.004}
& \textbf{0.246} / \textbf{+0.008}
& \textbf{0.061} / \textbf{+0.001}
& \textbf{0.061} / \textbf{-0.000}
& \textbf{0.060} / \textbf{+0.017}
& 0.062 / -0.002 \\

& Iso
& 0.253 / -0.020
& 0.251 / -0.012
& 0.257 / -0.035
& 0.255 / -0.025
& 0.062 / -0.002
& 0.062 / -0.001
& 0.062 / -0.001
& \textbf{0.061} / \textbf{+0.002} \\

\midrule


\multirow{3}{*}{DeepSeekCoder}
& Before
& 0.300 / -0.341
& 0.253 / -0.130
& 0.228 / -0.019
& 0.321 / -0.435
& 0.659 / -3.024
& 0.481 / -1.940
& 0.252 / -0.541
& 0.169 / -0.035 \\

& PS
& \textbf{0.224} / \textbf{+0.001}
& \textbf{0.222} / \textbf{+0.007}
& \textbf{0.223} / \textbf{+0.004}
& 0.222 / +0.008
& 0.164 / +0.000
& \textbf{0.163} / \textbf{+0.001}
& \textbf{0.163} / \textbf{+0.003}
& \textbf{0.163} / \textbf{+0.001} \\

& Iso
& 0.226 / -0.010
& 0.223 / +0.004
& 0.228 / -0.017
& \textbf{0.221} / \textbf{+0.013}
& \textbf{0.162} / \textbf{+0.010}
& 0.164 / -0.001
& 0.164 / -0.004
& 0.164 / -0.002 \\

\midrule


\multirow{3}{*}{OpenCoder}
& Before
& 0.250 / -0.219
& 0.213 / -0.039
& 0.236 / -0.149
& 0.378 / -0.840
& 0.666 / -2.667
& 0.532 / -1.931
& 0.305 / -0.677
& 0.200 / -0.099 \\

& PS
& \textbf{0.205} / \textbf{-0.000}
& \textbf{0.205} / \textbf{+0.001}
& 0.205 / -0.000
& \textbf{0.203} / \textbf{+0.008}
& \textbf{0.182} / \textbf{+0.000}
& \textbf{0.182} / \textbf{+0.000}
& \textbf{0.182} / \textbf{-0.000}
& \textbf{0.182} / \textbf{-0.001} \\

& Iso
& 0.210 / -0.025
& 0.205 / -0.000
& \textbf{0.205} / \textbf{+0.000}
& 0.209 / -0.016
& 0.183 / -0.006
& 0.183 / -0.005
& 0.182 / -0.003
& 0.183 / -0.006 \\

\midrule


\multirow{3}{*}{QwenCoder}
& Before
& 0.273 / -0.214
& 0.218 / +0.031
& 0.254 / -0.128
& 0.426 / -0.896
& 0.781 / -9.318
& 0.535 / -6.068
& 0.374 / -3.941
& 0.152 / -1.004 \\

& PS
& 0.223 / +0.010
& \textbf{0.217} / \textbf{+0.036}
& 0.219 / +0.024
& \textbf{0.219} / \textbf{+0.024}
& \textbf{0.075} / \textbf{+0.003}
& 0.076 / -0.003
& \textbf{0.074} / \textbf{+0.016}
& \textbf{0.076} / \textbf{+0.002} \\

& Iso
& \textbf{0.218} / \textbf{+0.029}
& 0.220 / +0.022
& \textbf{0.216} / \textbf{+0.041}
& 0.224 / +0.005
& 0.076 / +0.000
& \textbf{0.076} / \textbf{-0.003}
& 0.076 / -0.002
& 0.076 / -0.005 \\

\bottomrule
\end{tabular}%
}
\end{table*}

\subsection{Reliability Under Finite Sampling Budgets}

\textbf{Evaluation Setup.}
We next evaluate whether repeated sampling can compensate for overconfident single generations. We run four code models on MBPP+ and generate up to (K=50) candidate programs per task using temperature sampling with (T=0.8). Each candidate is executed against the task test suite. For each task, we record whether a passing solution appears within the first (k) samples and how many passing candidates appear within the full budget.

\begin{figure}[t]
  \centering
  \begin{subfigure}[t]{0.32\linewidth}
    \centering
    \includegraphics[width=\linewidth]{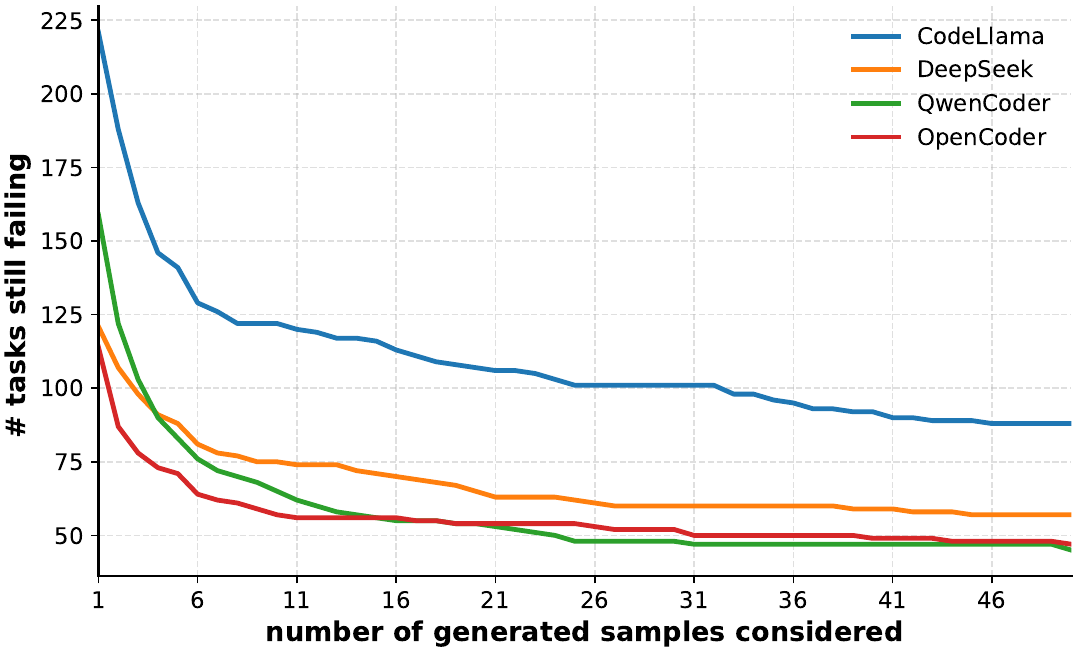}
    \caption{Tasks still failing vs. $k$.}
    \label{fig:plus_fail_curve_sub}
  \end{subfigure}
  \begin{subfigure}[t]{0.32\linewidth}
    \centering
    \includegraphics[width=\linewidth]{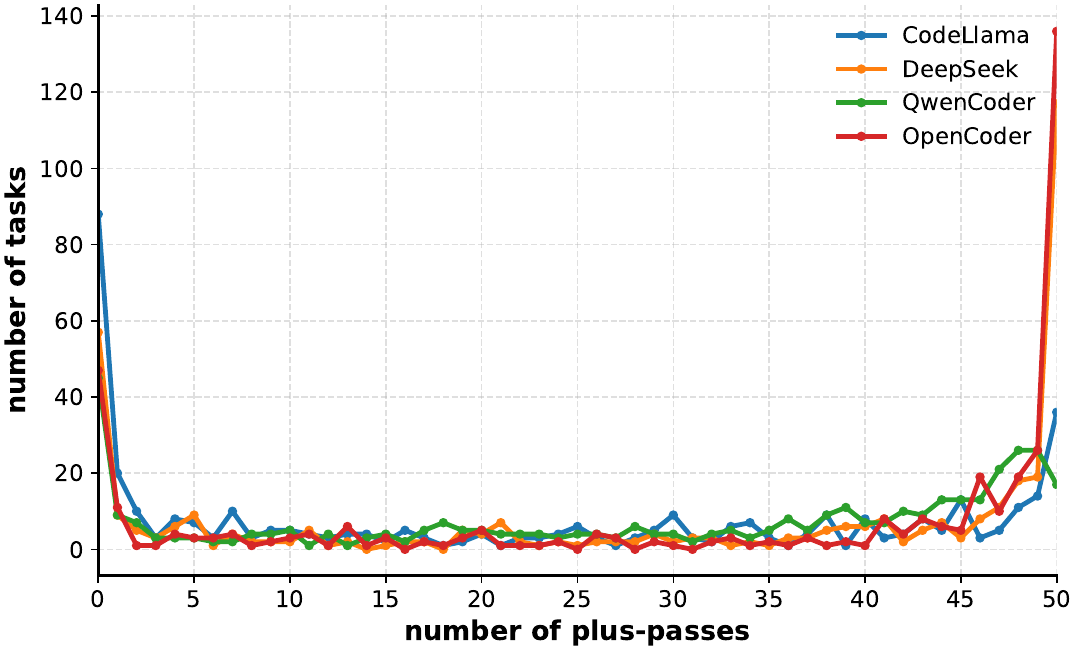}
    \caption{Per-task pass-count distribution ($K{=}50$).}
    \label{fig:plus_passcount_hist_sub}
  \end{subfigure}
  \caption{Sampling outcomes under EvalPlus+. (a) Remaining-unsolved tasks after the first $k$ generations. (b) Distribution of how many successes each task attains within $K{=}50$ generations.}
  \label{fig:evalplus_sampling_outcomes}
\end{figure}

\textbf{Results.}
Figure~\ref{fig:evalplus_sampling_outcomes} shows that sampling reduces the number of unsolved tasks quickly at small budgets, but the curves flatten as (K) increases. Even after 50 attempts, many tasks remain unsolved, i.e., roughly 90 for CodeLlama, about 57 for DeepSeek, and around 45--50 for OpenCoder and QwenCoder. Figure~\ref{fig:plus_passcount_hist_sub} shows that sampling mostly separates tasks into reliably solved and persistently unsolved groups, with fewer partially recoverable cases. 



These results characterize sampling as a bounded recovery mechanism rather than a reliability mechanism. Additional samples can recover failures that are sensitive to decoding variation, but the flattening curves show that many tasks remain unresolved even after substantial retry budget. In a software-engineering workflow, each retry also incurs generation cost, latency, and additional validation effort. Thus, sampling does not remove the accept-or-defer decision; it shifts the decision to budget allocation. A practical system must decide when another sample is likely to help, when a task should be escalated, and how to identify which candidate is trustworthy when multiple candidates are available.

\noindent\textbf{Finding.}
Sampling expands the candidate set but does not solve candidate selection or persistent failure.


\subsection{Explicit Abstention Prompting Does Not Reliably Induce Deferral}

\textbf{Evaluation Setup.}
We test whether explicit prompting can make a model defer when it is unlikely to be correct. We use a penalty-aware prompt that instructs the model to answer only when it is more than 95\% confident and to otherwise output \textbf{``I don't know''}. This tests whether a code LLM can convert uncertainty into abstention without additional control logic.



\begin{figure}[t]
    \centering

    \begin{minipage}[t]{0.56\columnwidth}
        \centering
        \includegraphics[width=\linewidth]{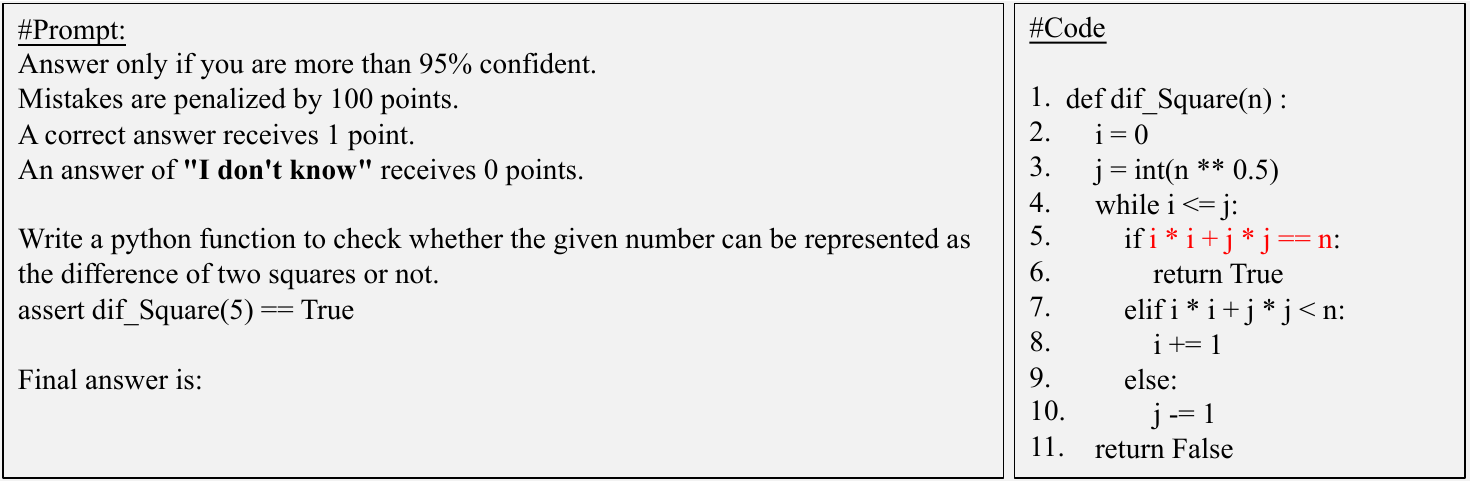}
        \captionof{figure}{
            Ablation: Explicit abstention prompting still yields
            incorrect code instead of deferral on MBPP/72 for OpenCoder.
        }
        \label{example}
    \end{minipage}
    \hfill
    \begin{minipage}[t]{0.43\columnwidth}
        \centering

        \begin{subfigure}[t]{0.49\linewidth}
            \centering
            \includegraphics[width=\linewidth]{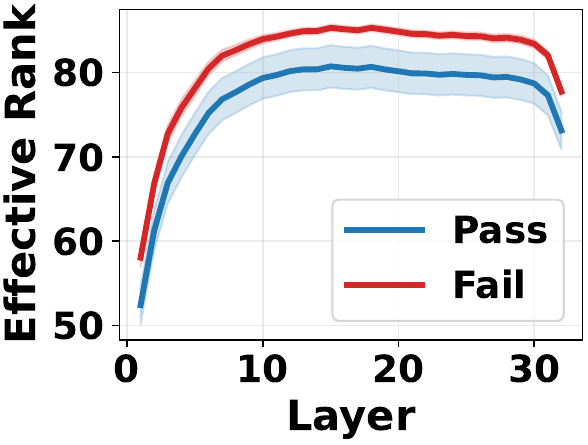}
            \caption{CodeLlama.}
            \label{fig:codellama}
        \end{subfigure}
        \begin{subfigure}[t]{0.49\linewidth}
            \centering
            \includegraphics[width=\linewidth]{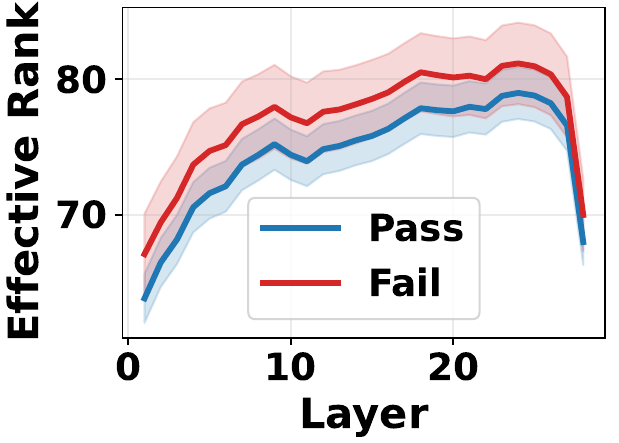}
            \caption{QwenCoder.}
            \label{fig:qwencode}
        \end{subfigure}

        \captionof{figure}{
            Layerwise effective rank for CodeLlama and QwenCoder
            on HumanEval+.
        }
        \label{fig:effective_rank}
    \end{minipage}

\end{figure}

\textbf{Results.}
Figure~\ref{example} shows a representative failure on MBPP/72 for OpenCoder. The prompt instructs the model to answer only when confident and assigns a high penalty to mistakes, while allowing \textbf{``I don't know''} as a zero-cost response. Nevertheless, the model generates code rather than deferring. The generated program checks whether the input can be represented as a sum of two squares (line 5), while the task asks for the difference of two squares. We observe the same qualitative pattern across the subject models: even when abstention is explicitly allowed,
the standard code-LLMs behavior favors producing a concrete program instead of deferring. In SE terms, the model continues to behave as a generator rather than as a reliability-aware component.


\noindent\textbf{Finding.}
Prompt-level abstention does not reliably induce deferral; models can still generate incorrect code even when an \textbf{IDK} option is provided.




\subsection{Implications}

Our results suggest that generation quality and correctness-sensitive confidence are distinct reliability properties. Improving the probability of generating correct code does not necessarily improve the separation between correct and incorrect generations in confidence. Overconfidence exposes this gap because incorrect generations can remain highly confident even when generation quality or probability alignment improves.

\textbf{SE Takeaway.}
Code-generation reliability should be evaluated along both dimensions. Pass rate measures whether a model produces correct programs, whereas overconfident failure captures whether incorrect programs remain highly trusted by the model's confidence signal. Improvement in generation quality should not be interpreted as a corresponding reduction in overconfidence.

\textbf{Research Insight.}
Overconfidence motivates treating correctness-sensitive confidence as an explicit learning objective rather than an expected consequence of better code generation. Future techniques should aim to reduce confidence on incorrect generations while preserving confidence on correct generations. This formulation extends the reliability objective from producing correct code to distinguishing which generated code can be trusted.

\section{Latent-Space Evidence of Overconfident Failure}
\label{rq4}

This section addresses \textbf{RQ4} by investigating overconfident failure signals in latent representation-space of code-LLMs. 

\textbf{Evaluation Setup.}
Inspired by recent latent-space analyses of reasoning models~\cite{du2025latent}, we analyze the latent representation space across transformer layers. For each generated solution, we feed the prompt and generated code to the model and compute hidden-state statistics only over the generated solution tokens. We focus on \textit{effective rank}~\cite{wei2024diff}, which summarizes how many representation directions are effectively used at a layer. Given a hidden-state matrix ($H_\ell$) at layer ($\ell$), effective rank is computed from the normalized singular-value spectrum of ($H_\ell$). Higher effective rank indicates a more dispersed representation, while lower effective rank indicates a more compact representation.

\textbf{Results.}
Figure~\ref{fig:effective_rank} compares layerwise effective rank for CodeLlama and QwenCoder on HumanEval+. Although the models operate at different pass@1 levels, both show the same pattern: failing generations have higher effective rank than passing generations across many middle and later layers. This indicates that failing generations induce a less compact hidden-state geometry. This contrasts with RQ2, where output-token confidence often failed to separate passing from failing programs. The result suggests that correctness-related variation may be partially encoded internally but not exposed through token probabilities.

\noindent\textbf{Finding.}
Passing and failing generations differ in latent representation-space geometry even when output confidence remains unreliable.


\noindent\textbf{SE Takeaway.}
A confident decoding trace should not be interpreted as a low-risk program. Representation-level signals may help future tools identify candidates that appear reliable from token probabilities but still require validation.

\noindent\textbf{Research Insight.}
The effective-rank separation suggests a representation-aware direction for mitigating overconfident failure. If failing generations occupy a different latent-state regime, then correctness risk is partially represented internally but not coupled to decoding behavior. Future SE research should therefore study such objectives that make these internal risk signals operational: lowering confidence, triggering deferral, or routing candidates to validation when hidden-state geometry indicates likely failure. This shifts uncertainty from an external post-hoc score to a learned mapping from representations to reliability actions.

\section{Discussion}
\label{sec}

Our findings suggest three directions for reducing overconfident failure: training, inference-time control, and reliability evaluation.

\textbf{1) Train models to recognize likely failures.}
Our post-training results show that better generation does not imply better discrimination between passing and failing programs. Instruction tuning increases certainty on failing generations for both model families, while its effect on pass--fail discrimination is inconsistent. Therefore, future post-training should optimize not only whether the model generates correct code, but also whether it lowers confidence or defers when its generated code is likely incorrect.

This direction requires execution-grounded training data. Each training example should connect a task and generated candidate to an execution outcome, so that high-confidence failures can be identified explicitly. Matched passing and failing candidates for the same task are particularly useful because they control for differences in task difficulty. Such data can support objectives that penalize confident failures and reward appropriate deferral. Our latent-space analysis provides an additional opportunity where passing and failing generations exhibit different internal representations even when output confidence overlaps, suggesting that these internal signals may support learned failure prediction rather than relying only on token probabilities. 

\textbf{2) Couple model confidence with external evidence at inference time.}
Our selective-generation results show that high confidence alone is not sufficient for acceptance. In some settings, the most certain generations contain disproportionately many failures. Therefore, inference-time systems should not map confidence directly to ``accept'' or ``defer.'' Instead, confidence should be one input to a control policy that determines what additional evidence is needed.


At inference time, the system should determine what additional evidence is needed before accepting generated code. We distinguish three such cases. First, when the task specification is incomplete or ambiguous, the system should seek clarification from the developer. Second, when correctness depends on missing external information, the system should retrieve or verify that information. Third, when the risk lies in the generated implementation, the system should obtain program-level evidence through testing, analysis, or an independent evaluator. Repeated generation can support this process, but our results show that sampling alone does not resolve candidate selection and eventually reaches persistent failures. Therefore, an inference-time controller should choose among clarification, validation, regeneration, and deferral according to the expected reliability improvement and its cost.

\textbf{3) Evaluate reliability through system decisions.}
Our results show that score-level improvements can overstate practical reliability. Calibration substantially reduces BS while BSS remains near zero in many settings, and explicit IDK prompting does not reliably induce deferral. Thus, future evaluation should assess whether uncertainty leads to better acceptance and validation decisions rather than only whether a confidence score is numerically well aligned. 

A reliability benchmark should evaluate an end-to-end policy that can accept generated code, obtain additional validation, retry generation, or defer. The evaluation should vary the cost of accepting incorrect code and the available validation budget. The central outcome is whether the system reduces false acceptance without relying on indiscriminate deferral or unlimited validation. Such evaluation would measure whether a code-generation system uses reliability evidence appropriately under deployment constraints, rather than only whether it produces a calibrated score.

These three directions target different causes of overconfident failure. Training determines whether the model identifies likely failures, inference-time control determines how the system responds to uncertain evidence, and evaluation determines whether those responses reduce false acceptance. Ultimately, an intervention addresses overconfidence only to the extent that it reduces failing programs that remain eligible for acceptance, without relying on indiscriminate deferral or unbounded validation.

\section{Limitations}
\label{sec:limitations}

Our study has three main limitations. First, we evaluate four open-source code LLMs in the 7--8B range because our analyses require token probabilities and hidden representations and whether the findings extend to larger proprietary models remains open. Second, we study single-generation code synthesis rather than repository-scale or agentic workflows, where tool use and iterative feedback may change how overconfidence emerges. Third, our evaluation is limited to Python benchmarks with execution-based test suites. These provide practical correctness labels but do not establish complete semantic correctness or cover other languages and development settings.

\section{Threats to Validity}
\label{threats}

This section discusses potential threats to validity in our study, organized into four standard categories: internal, external, construct, and conclusion validity. These categories address risks pertaining to our experimental design and methodology, as well as the generalizability of our findings.

\textbf{Internal validity.}
Our results depend on the correct implementation of uncertainty metrics, calibration methods, sampling procedures, and execution-based evaluation. We mitigate this threat by following standard definitions from prior work~\cite{vashurin2024benchmarking, fomicheva2020unsupervised, fadeeva2024fact, takayama2019relevant, van2022mutual, darrin2022rainproof, duan2023shifting, kuhn2023semantic, lin2023generating}, reusing established implementations when available, and applying the same evaluation pipeline across models and benchmarks. For stochastic procedures, such as sampling-based uncertainty and repeated generation, we use fixed decoding settings within each experiment and compare trends across multiple models. Nevertheless, implementation details such as token filtering, sequence truncation, and score orientation may affect individual metric values.

\textbf{External validity.}
We evaluate four open-source code LLMs on HumanEval+, MBPP+, and BigCodeBench analyses. These benchmarks provide executable correctness labels and allow controlled comparison of uncertainty signals, but they do not cover all software-engineering settings. Our findings may not fully generalize to larger proprietary models, project-level development tasks, other programming languages, or agentic workflows with tool use and iterative repair. Metrics requiring token probabilities or hidden states also assume white-box model access, which may not be available for API-only models. Extending this analysis to closed models and repository-scale tasks remains future work.

\textbf{Construct validity.}
We operationalize correctness using benchmark test suites. Although HumanEval+ and MBPP+ strengthen the original tests, execution outcomes may still miss untested behaviors. Thus, our labels approximate functional correctness rather than prove semantic equivalence. We also treat token confidence, entropy, calibration scores, sampling behavior, abstention, and latent effective rank as proxies for reliability-related evidence. These constructs do not directly measure program correctness; they measure whether model-derived signals align with execution outcomes. This distinction is central to our study, but it limits claims about absolute correctness.

\textbf{Conclusion validity.}
Our conclusions are based on trends observed across multiple models, benchmarks, uncertainty families, and mitigation strategies. The main findings do not depend on a single metric: weak failure detection, high-confidence failures, limited local-token separation, bounded sampling gains, and unreliable prompt-level abstention appear across settings. However, exact effect sizes may vary with decoding parameters, calibration splits, thresholds, and benchmark composition. We therefore avoid claiming that uncertainty is useless. Our conclusion is more specific: the evaluated model-derived signals are insufficient as standalone reliability mechanisms for deciding that generated code is likely to pass execution-based validation.

\section{Related Work}
\label{related}

\textbf{Uncertainty estimation in ML and NLG.}
Uncertainty estimation has been widely studied in classification, using confidence scores, entropy, Bayesian approximations, ensembles, and calibration methods to identify unreliable predictions~\cite{guo2017calibration,wang2020dissector,hendrycks2018baseline,gal2016dropout,alon2019code2vec,xiao2019quantifying,vasudevan2019towards,corbiere2019addressing,monarch2021human,steinhardt2016unsupervised,shannon1948mathematical}. Recent work extends these ideas to natural-language generation through token uncertainty, sequence likelihood, sample diversity, semantic uncertainty, and judge-based estimates~\cite{vashurin2024benchmarking,xie2025empirical,xia2025survey,bakman2025reconsidering,ielanskyi2025addressing,huang2025look,hu2024towards,li2024investigating,hu2023uncertainty,baan2023uncertainty,niwa2024ambignlg,kuhn2023semantic}. Code generation differs because correctness is determined by executable program behavior rather than textual plausibility or subjective quality. Our work studies whether these uncertainty signals remain useful under execution-based correctness.

\textbf{Uncertainty and calibration for code models.}
Prior work has studied uncertainty, calibration, and reliability signals for code models~\cite{li2024doce,peng2025perfcodegen,ni2023lever,tian2025fixing,huang2025look,vasconcelos2022generation,spiess2024calibration,zhou2024calibration,sharma2025assessing,pmlrjohnson23a,rathnasuriya2026answer, shi2026code,rakasi2026uncertainty,johnson2023ru,rathnasuriya2025codeimprove}. These studies show that code models can be miscalibrated and that post-hoc methods can improve confidence alignment. Our focus is different: we study whether model-derived uncertainty can support reliability decisions for generated code, such as identifying likely failures, abstaining, or deciding when additional validation is needed. This setting is distinct from uncertainty in non-generative SE tasks such as defect prediction or vulnerability detection~\cite{li2021estimating,rathnasuriya2025codeimprove}, where predictions are labels over existing artifacts rather than generated programs whose correctness is observed only after execution.

\textbf{Code-generation evaluation and mitigation.}
Code-generation research commonly evaluates models with pass@1, pass@k, execution feedback, repair, and prompting strategies~\cite{li2024doce,peng2025perfcodegen,ni2023lever,tian2025fixing,zheng2024opencodeinterpreter,liu2024exploring}. These methods measure or improve the ability to produce correct code, but they do not fully address whether a model can recognize when its own output is likely to fail. Sampling can increase the chance that a correct candidate exists, but does not identify which candidate should be trusted. Prompt-level abstention can expose a deferral option, but does not guarantee safe generate-or-defer behavior. We evaluate these techniques as responses to overconfident failure rather than as general performance-improvement methods.

\textbf{Summary.}
Our work provides a rigorous characterization of the \emph{dilemma of overconfidence} in code LLMs: model-derived uncertainty is available during generation, but it does not reliably identify generated programs that fail execution-based evaluation. 
The results position overconfident failure as a deployment-facing SE reliability problem and motivate future mechanisms that connect model self-assessment to execution-grounded correctness.

\section{Conclusion}
\label{conclusion}
This paper characterizes the \emph{dilemma of overconfidence} in execution-grounded code generation. We show that model-derived uncertainty often fails to identify incorrect programs, even under calibration, sampling, and abstention prompting. These results motivate behavior-aware evaluation that measures high-confidence failures, generate-or-defer behavior, and reliability under limited validation budgets.

\clearpage
\bibliographystyle{IEEEtran}
\bibliography{hatemoderate}
\end{document}